\documentclass[%
 reprint,
superscriptaddress,
nofootinbib,
 amsmath,amssymb,
 aps,
prd,
]{revtex4-1}

\usepackage[usenames, dvipsnames, table]{xcolor}	%
\usepackage[normalem]{ulem}					%
\usepackage{enumitem}						%
\usepackage{setspace}						%
\usepackage{caption}						%
\usepackage{subcaption}						%
\usepackage[section]{placeins}					%
\usepackage{hyperref}						%
\hypersetup{
    colorlinks,
    linkcolor={blue!50!black},
    citecolor={blue!50!black},
    urlcolor={blue!80!black}
}		%
\usepackage{xurl}							%
\usepackage{multirow}						%
\usepackage{hhline}							%
\usepackage{afterpage}						%
\usepackage{fancyhdr}						%
\usepackage{pifont}							%

\usepackage{braket}					%
\usepackage{bbm}					%
\usepackage{colonequals}				%
\usepackage{tensor}					%
\usepackage[scr=boondoxo]{mathalfa}	%
\usepackage{extpfeil}				%
\usepackage{qcircuit}				%
\usepackage{simplewick}				%

\newcommand{\al}{\alpha}
\newcommand{\be}{\beta}
\newcommand{\ga}{\gamma}

\newcommand{\tht}{\theta}	%

\newcommand{\la}{\lambda}
\newcommand{\rh}{\rho}
\newcommand{\si}{\sigma}
\newcommand{\ta}{\tau}

\newcommand{\ph}{\phi}

\newcommand{\De}{\Delta}

\newcommand{\Ph}{\Phi}
\newcommand{\Ps}{\Psi}
\newcommand{\Om}{\Omega}

\newcommand{\cellcolour}{\cellcolour}			%
\newcommand{\eqn}[1]{\begin{equation}#1\end{equation}}				%
\newcommand{\aln}[1]{\begin{align}#1\end{align}}					%
\newcommand{\pmx}[1]{\begin{pmatrix}#1\end{pmatrix}}	%
\newcommand{\lb}{\left}						%
\newcommand{\rb}{\right}					%
\newcommand{\nn}{\nonumber}					%
\newcommand{\tx}{\text}						%
\newcommand{\mc}{\mathcal}					%
\newcommand{\Or}{\mathcal{O}}				%
\newcommand{\bb}{\mathbb}					%
\newcommand{\ce}{\colonequals}				%
\newcommand{\tp}{\otimes}					%
\newcommand{\ct}{\dagger}					%
\newcommand{\infi}{\infty}					%
\newcommand{\ti}{\widetilde}				%
\newcommand{\Bk}{\Braket}					%
\newcommand{\iu}{\mathrm{i}}		%
\newcommand{\ec}{\mathrm{e}}	%
\DeclareMathOperator{\Tr}{Tr}				%
\DeclareMathOperator{\arccosh}{arccosh}		%
\newcommand{\siy}{\hat{\sigma}_y}		%
\newcommand{\Abs}[1]{\left|{#1}\right|}		%
\newcommand{\Kb}[2]{\Ket{#1}\Bra{#2}}		%
\makeatletter
	\newcommand{\vast}{\bBigg@{3}}
	\newcommand{\Vast}{\bBigg@{4}}
\makeatother

\renewcommand{\Re}{\operatorname{Re}}

\usepackage{graphicx}
\graphicspath{{figures/},{pictures/},{NewFigs/}}

\begin{document}

\title{Entanglement Harvesting with a Twist}
\author{Laura J. Henderson}
\affiliation{Department of Physics and Astronomy, University of Waterloo, Waterloo ON, Canada, N2L 3G1} 
\affiliation{Institute for Quantum Computing, University of Waterloo, Waterloo ON, Canada, N2L 3G1}
\author{Su Yu Ding}
\affiliation{Department of Physics and Astronomy, University of Waterloo, Waterloo ON, Canada, N2L 3G1} 
\author{Robert B. Mann}
\affiliation{Department of Physics and Astronomy, University of Waterloo, Waterloo ON, Canada, N2L 3G1}
\affiliation{Institute for Quantum Computing, University of Waterloo, Waterloo ON, Canada, N2L 3G1}
\affiliation{Perimeter Institute for Theoretical Physics, 31 Caroline Street North, Waterloo ON, Canada, N2L 2Y5}
\affiliation{Waterloo Centre for Astrophysics, University of Waterloo, Waterloo ON, Canada, N2L 3G1}

\begin{abstract}
    One consequence of the cosmic censorship conjecture is that any  topological structure  will ultimately collapse to within the horizons of a set of black holes, and as a result, an external classical observer  will be unable to probe it.
    However a single two-level quantum system (UDW detector) that remains outside of the horizon has been shown to distinguish between a black hole and its associated  geon counterpart via its different response rates.  Here  we extend
this investigation of the quantum vacuum outside of an $\bb{RP}^2$ geon  by considering the entanglement structure of the vacuum state of a quantum scalar field in this spacetime, and how this differs from its BTZ black hole counterpart. Employing the entanglement harvesting protocol, where field entanglement is swapped to a pair of UDW detectors,   we find that 
the classically hidden topology of the geon can have 
an appreciable difference in the amount of entanglement harvested in the two spacetimes 
for sufficiently small  mass.   In this regime, we find that detectors with a small energy gap harvest more entanglement in the BTZ spacetime; however as the  energy gap increases, the detectors harvest more entanglement in a geon spacetime.  The energy gap at the crossover is dependent on the black hole mass,  occurring at a lower values for lower masses.  This also impacts the size of the entanglement shadow, the region near the horizon where the detectors cannot harvest entanglement.  Small gap detectors experience a larger entanglement shadow  in a geon spacetime, whereas for large gap detectors the shadow  is larger in a BTZ spacetime.  
\end{abstract} 

\maketitle

\section{Introduction}
\label{Introduction}

It is not at all an exaggeration to say that the introduction of topology into physics was inaugurated by Sir Roger Penrose.  
While at the 1954 International Conference of Mathematicians in Amsterdam as a student, he  was inspired by an exhibition of the work of Dutch artist M.C. Escher, and soon began to conceptualize impossible objects of his own \cite{Welch2012,Kumar2010}.  One of these is known as the Penrose triangle, or tribar -- a shape  that looks like a solid three dimensional triangle, but is not.  With his father Lionel, he developed the Penrose staircase, a set of stairs that one could perpetually climb (or descend) in an impossible loop.  These ideas in turn inspired Escher to produce two of his most famous masterpieces, {\it Waterfall} and {\it Ascending and Descending} \cite{AandD2020}.

It is fortunate for physics -- indeed for science -- that Penrose turned his attention from mathematics to astrophysics, introducing mathematical tools that forever changed how we understand and analyze spacetime.  His innovative perspective was to concentrate on the topology of spacetime instead of its detailed geometric structure.  This led to the notion of the conformal structure of spacetime, which determines the trajectories of null geodesics, and in turn the causal relationships between different regions of spacetime.

His groundbreaking paper on gravitational collapse \cite{PhysRevLett.14.57} demonstrated (in the context of classical physics) that ultimately some kind of spacetime singularity will form from  imploding matter (for example a collapsing star) provided energy is positive and the field equations of general relativity hold.  This in turn led to the notion of cosmic censorship, Penrose's 1969 conjecture 
\cite{Penrose:1969,Penrose:1999vj} that any singularities formed in a physical process will be confined within a well-behaved event horizon that surrounds a hidden space-time region, from which the term black hole was coined by Wheeler \cite{WheelerBH:2000}. 

The cosmic censorship conjecture implies that an event horizon will hide any singularities forming to the future of a regular initial data surface.  If this Cauchy surface is not simply connected (and asymptotically flat), it will also have singular time evolution provided the weak energy condition holds \cite{Gannon:1975}.    Presumably any topological structures will ultimately collapse within the horizon of a  black hole (or a collection of them) if cosmic censorship holds.  This implies that there is a topological  censorship conjecture: no (classical) observer remaining outside a black hole can probe the topology of spacetime \cite{Friedmann:1991}, which has been proven for globally hyperbolic spacetimes  \cite{Friedman:1993ty}.  Essentially, any topological structure collapses too rapidly for light to traverse it.

Such theorems do not prevent the existence of spacetimes that have non-simply connected structures hidden behind event horizons.  An interesting class of such objects are geons  \cite{Friedman:1993ty,Louko:1998dj,Louko:2004ej}.  A simple example is the 
$\mathbb{RP}^3$ geon:  a  space and time orientable $\mathbb{Z}_2$ quotient of the Kruskal manifold. It has  an asymptotically flat exterior region that isometric to a standard Schwarzschild exterior and  hence is an eternal black hole.     It is possible to generalize such objects to have spin and charge \cite{Louko:2004ej}, and to have other asymptotic structures apart from flatness.

Classically, to observers outside the event horizon, geons are indistinguishable from their standard  black hole counterparts.  Indeed,
as an eternal black and white hole spacetime, we don't expect one to form by some astrophysical process.  What warrants their consideration in physics is precisely their status as   unconventional black holes, particularly in quantum contexts.  The $\mathbb{RP}^3$ geon was not only used to illustrate cosmic censorship, but was also shown to raise interesting questions about black hole entropy and its statistical mechanical interpretation in the context of the quantum vaccum for scalar and spinor fields  \cite{Louko:1998dj}.

 The $\mathbb{RP}^2$ geon -- a  (2 + 1) dimensional asymptotically anti de Sitter (AdS) analogue of the $\mathbb{RP}^3$ geon  --
 has been a particularly interesting object to study.  Its black hole counterpart is the non-rotating Banados-Teitelboim-Zanelli (BTZ) black hole
  \cite{Banados:1992,Banados:1993}, which is an interesting theoretical laboratory for studying quantum gravity \cite{Carlip:2003}.  In particular, the `BTZ geon'
  has been used to probe AdS/CFT correspondence \cite{Louko:1998hc,Louko:2000tp,Maldacena:2001kr}.   The boundary 
 Conformal Field Theory (CFT)  state  corresponding to the geon is pure, but has correlations that yield  thermal expectation values at the usual BTZ Hawking temperature (for suitably restricted classes of operators), suggesting an interesting relationship between topology and correlations in the boundary state.   The geon has also been shown to have interesting duality  properties relating  the volume of  a maximum time slice and a   quantum information metric in the dual CFT \cite{Sinamuli:2016rms,Sinamuli:2018jhm}
 that are natural counterparts of the BTZ case \cite{Miyaji:2015woj}, again indicating that hidden topology can manifest itself in other physical effects.
 
In fact quantum effects can be used to actually  probe the interior of a geon, allowing one to `look' inside a black hole \cite{Smith:2014,Ng2017}.
In particular, the transition rates of an Unruh-DeWitt (UDW) detector \cite{Unruh1976,deWitt} placed outside the horizon of an eternal BTZ black hole and its associated geon counterpart differ.  The geon case is   time-dependent, implying  the topological structure of the singularity  can indeed be probed from outside the Killing horizon.
  
All of this makes it  clear that the quantum vacuum outside a geon has a considerably different character than that outside of a black hole.   Here we are concerned with probing the entangling properties of spacetime outside such objects to see how they differ from
those of their black hole counterparts.  Specifically, we study the vacuum entanglement that two UDW detectors can extract from spacetime outside of the $\mathbb{RP}^2$ geon as compared to its previously studied 
\cite{henderson:2018,AH:plb2020,Robbins:2020jca} BTZ counterpart.  We find that the censored topology of the geon does indeed affect the vacuum entanglement properties of a scalar field, manifest as differences in non-local correlations between the detectors and in the amount of entanglement they can extract.

The entanglement properties of quantum field theories, in particular the entanglement present in the vacuum, has  applications in multiple areas of physics including quantum information  \cite{Peres2004,Lamata1997}  and metrology \cite{Ralph2009}, the AdS/CFT correspondence \cite{Ryu2006},   quantum energy teleportation~\cite{doi:10.1143/JPSJ.78.034001,Hotta:2011xj}, and black hole entropy \cite{Solodukhin2011,Brustein2005}.  Indeed they are at the   core 
the black hole information paradox~\cite{Preskill:1992tc,Mathur:2009hf} and its proposed solutions~\cite{Almheiri:2012rt,Braunstein:2009my,Mann:2015luq}.

It has  long been known~\cite{summers1985bell,summers_bells_1987}  that quantum vacuum correlations
are present between   both timelike and spacelike separated regions.   Several years later it was   realized   \cite{Valentini1991} that    this vacuum entanglement can be swapped with 
a physical system: two initially uncorrelated atoms (either spacelike or timelike separated) that interact for a finite time  with the electromagnetic vacuum can exhibit nonlocal correlations. 

Understanding this process is best   explicated by using UDW detectors to model the atoms, idealized as two-level qubits, 
and a scalar field to model the electromagnetic one \cite{Reznik2003,Reznik2005}.  Entanglement in the scalar quantum vacuum  can be transferred to the UDW detectors via a protocol known as entanglement harvesting  \cite{Salton2015}, and is particularly 
useful insofar as it   characterizes properties of the quantum vacuum inaccessible to a single detector.  Examples include probes of spacetime topology \cite{Smith2016},  the distinctive thermal character of de Sitter spacetime 
\cite{PhysRevD.79.044027,Huang2017},   and the discovery of new structures such as separability islands in anti-de Sitter 
spacetime \cite{Ng:2018drz,Henderson2019}. 

Over the past few years we have been learning more about  harvesting vacuum entanglement  in black hole spacetimes. 
  The original investigation    revealed that a black hole has an entanglement shadow: a region outside its horizon within which it is not possible to  harvest entanglement \cite{henderson:2018}.   A more recent study \cite{Robbins:2020jca} showed that the situation dramatically changes if the black hole is rotating: the shadow persists, but the extracted entanglement can be amplified by as much as a factor of 10 at moderate distances away from the horizon for near-extremal black holes.   These studies were carried out for the  BTZ black hole \cite{Banados1992} but are expected to be
 universal \cite{Henderson:2019uqo}, and a recent study   in Schwarzschild/Vaidya spacetimes of entanglement harvesting  is consistent with this expectation \cite{Tjoa2020}.
 
Here we further advance our understanding of entanglement harvesting by investigating it outside an
$\mathbb{RP}^2$ geon  (or simply `geon', as we shall now refer to this object).  While large-mass geons have entangling properties that are effectively indistinguishable from their BTZ counterparts, small-mass geons exhibit quantitative distinctions, distinctions that increase as the mass gets smaller.  Non-local correlations are notably amplified in the small-mass geon case, though because the local noise (transition probability) of the detectors can likewise be amplified, this does not necessarily translate into a correspondingly larger amount of extractable entanglement.  

We shall begin by reviewing the entanglement harvesting protocol in section~\ref{sec:Entanglement harvesting in curved spacetimes} followed by a review of the geon spacetime and its BTZ counterpart.  We then compute in section~\ref{sec:EHDGS} the transition probability of a UDW detector 
outside of a geon, extending an investigation previously carried out for the transition rate \cite{Smith:2014}. We then compute the
non-local correlations of a pair of identical detectors, and from this determine the amount of entanglement they can extract. Our measure of entanglement is given by the concurrence  \cite{Smith2016,Smith:2017vle,Wooters1998}, and we examine its dependence on the separation between the horizon and the closest detector to it, as well as on the separation between the detectors and their identical energy gap.  Entanglement harvesting proves to be quite sensitive to this  latter quantity, which plays a significant role in
the reach of the entanglement shadow of the geon.  We close in section~\ref{sec:concl} with some conclusions on our work, and the relationship between quantum entanglement and spacetime topology.

\section{The entanglement harvesting protocol}
\label{sec:Entanglement harvesting in curved spacetimes}

\subsection{The Unruh-DeWitt detector}

We consider two identical two-level Unruh-DeWitt (UDW) detectors \cite{Unruh1976,deWitt} $A$ and $B$, with ground and excited states given by $\Ket{g}_D$ and $\Ket{e}_D$ respectively, separated by an energy gap $\Om_D$, which couples locally to a massless quantum scalar field $\hat{\ph}(x,t)$ and $D\in\{A,B\}$.  The trajectories of the detector are given by $x_D(\ta_D)$, where $\ta_D$ is the proper time of detector $D$.  The interaction of each detector is described buy the Hamiltonian
\begin{equation}
	H_D(\tau) = \lambda \chi_D(\tau) \Big( e^{i\Omega\tau}\sigma_D^+ + e^{-i\Omega\tau}\sigma_D^-\Big) \otimes \phi[x_D(\tau)],
\label{eq:int-Hamiltonian}
\end{equation}
where $\lambda\sqrt{\sigma} \ll1$ is the coupling strength of the interaction, $\chi_D(\tau)$ is the switching function, with characteristic width $\sigma$, that controls when the detector couples to the field, and the ladder operators associated with the Hilbert spaces of the detectors $\sigma^\pm$ are given by $\sigma_D^+ := \ket{1}_D\bra{0}_D$ and $\sigma_D^- := \ket{0}_D\bra{1}_D$. This simple model was shown to capture relevant features of the light-matter interaction when no angular momentum exchange is involved \cite{Martin-Martinez2013,Alvaro,Pozas-Kerstjens:2016}.

The evolution of the detector-field system, with respect to time $t$, is described by the unitary operator
\eqn{
    \hat{U} \ce \mc{T}\exp\lb[-\iu\int dt\lb(\frac{d\ta_A}{dt}\hat{H}_A[\ta_A(t)]+\frac{d\ta_B}{dt}\hat{H}_B[\ta_B(t)]\rb)\rb]
}
where $\mc{T}$ is the time ordering operator.

The detectors are initialized (at $t\to\-\infi$) in the ground state, and the field begins in the vacuum state, so the initial detector-field state is given by
\eqn{
    \Ket{\Ps_0} = \Ket{g}_A\tp\Ket{g}_B\tp\Ket{0}_\ph.
}
After the interaction (at $t\to\infi$), the density matrix describing the state of the two detector sub-system is given by $\hat{\rh}_{AB}=\Tr_{\ph}\lb[\hat{U}\Kb{\Ps_0}{\Ps_0}\hat{U}^\ct\rb]$, which is
\aln{
    \hat{\rh}_{AB} %
    &=\pmx{
        1-P_A-P_B & 0 & 0 & X^* \\
        0 & P_B & 0 & C^* & 0 \\
        0 & C & P_A & 0 \\
        X & 0 & 0 & 0
    }+\Or(\la^4)
    \label{eq:genrhoAB}
}
to lowest order in the coupling strength, where
\begin{widetext}
\aln{
	P_D &\ce \lambda^2 \int d\tau_D d\tau_D'\ \chi_D(\tau_D)\chi_D(\tau_D')
    \ec^{-\iu \Omega(\tau_D-\tau_D')} W\big(x_D(t),x_D(t')\big)
    \quad \text{for } D\in\{A,B\} \label{eq:general-P}\\
	C &\ce \lambda^2 \int d\tau_A d\tau_B\ \chi_A(\tau_A) \chi_B(\tau_B)
    \ec^{-\iu(\Omega_A\tau_A - \Omega_B\tau_B)} W\big(x_A(t),x_B(t')\big) \label{eq:general-C} \\
	X &\ce \lambda^2 \int d\tau_A d\tau_B\ \chi_A(\tau_A) \chi_B(\tau_B)
    \ec^{-\iu(\Omega_A\tau_A + \Omega_B\tau_B)}
    \Big[ \theta(t'-t)W\big(x_A(t),x_B(t')\big) + \theta(t-t')W\big(x_B(t'),x_A(t)\big) \Big]
\label{eq:general-X}
}
\end{widetext}
and $\tht(t-t')$ is the Heaviside theta function.  The function $W(x,x') = \Bk{0|\hat{\ph}\big(x(t),t\big),\hat{\ph}\big(x'(t'),t'\big)|0}$ is the Wightman function, the two-point correlator of the field between spacetime points $x(t)$ and $x'(t')$.  %

If either detector $A$ or detector $B$ is traced out of Eq.\ \ref{eq:genrhoAB}, the density matrix describing the final state of the remaining detector is
\eqn{
    \hat{\rh}_D = \pmx{
        1-P_D & 0\\
        0 & P_D
    } \quad \text{for } D\in\{A,B\}
}
so we can interpret $P_D$ as the transition probability of detector $D$.  The matrix elements $X$ and $C$ can be interpreted as the non-local correlations and the total correlations between the detectors respectively. %

\subsection{Quantifying the entanglement in $\rho_{AB}$}

For a pair of qubits, the concurrence is an entanglement monotone  that  ranges from a value of $0$ to $1$, and so we choose it as our measure of entanglement \cite{Wooters1998}.  It is defined by
\eqn{
    \mc{C}(\hat{\rh}) \ce \max\lb[0,w_1-w_2-w_3-w_4\rb]
}
where the $w_i$'s are the square roots of the eigenvalues $(w_1 \ge w_2 \ge w_3 \ge w_4)$ of the matrix 
\eqn{
    \hat{\rh}\lb[\lb(\siy\tp\siy\rb)\hat{\rh}^*\lb(\siy\tp\siy\rb)\rb]
}
and $\siy$ is the Pauli $y$ matrix.  For a density matrix of the form of Eq.\ \eqref{eq:genrhoAB}, the concurrence becomes
\eqn{
    \mc{C}(\hat{\rh}_{AB}) = 2\max\lb[0,\Abs{X}-\sqrt{P_AP_B}\rb].
}
This expression of concurrence provides a nice physical interpretation: detectors $A$ and $B$ are entangled when the non-local correlations $|X|$ dominates the root mean square of the local noise of the detectors, $P_A$ and $P_B$.

\subsection{The BTZ and Geon Spacetimes}

\begin{figure}
\includegraphics[width=\linewidth]{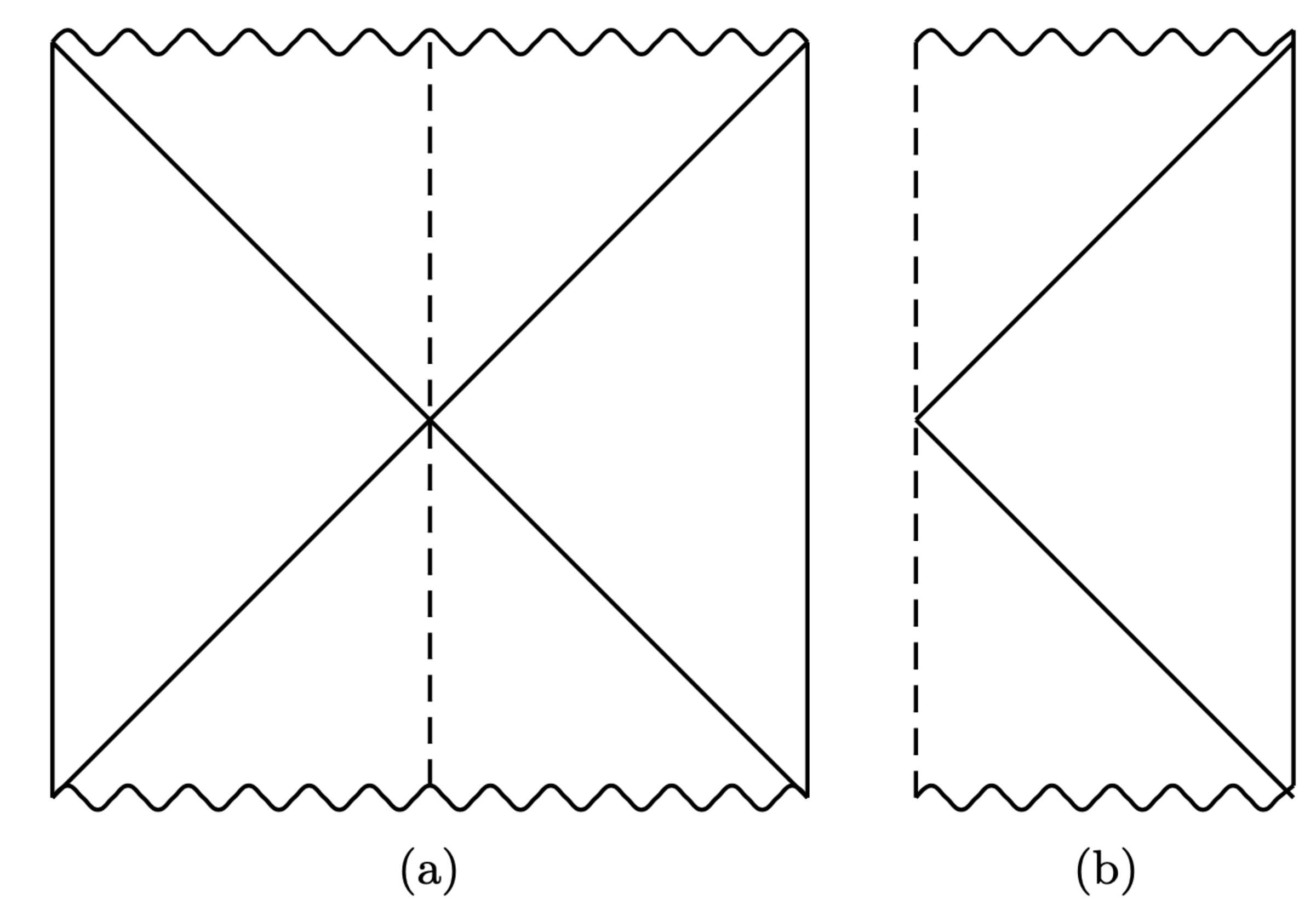}
\caption{%
     (a) A conformal diagram of the BTZ black hole;  each point in the diagram represents a suppressed $S^1$. (b) A conformal diagram of the $\mathbb{RP}^2$ geon spacetime; the region not on the dashed line is identical to the diagram in (a). However on the dashed line each point in the diagram again represents a suppressed $S^1$ but with half the circumference of the suppressed $S^1$ in diagram (a).
}
\label{fig:GeonConf}
\end{figure}

The BTZ black hole metric is a solution to Einstein's field equations in (2+1)-dimensions with a negative cosmological constant $\Lambda = -1/\ell^2$ \cite{Banados:1992,Banados:1993}. For a black hole with zero angular momentum, it is given by
\begin{eqnarray}
ds^2 &=&  -\Bigg(\frac{r^2-r_h^2}{\ell^2}\Bigg) dt^2 + \Bigg(\frac{r^2-r_h^2}{\ell^2}\Bigg)^{-1} dr^2 + r^2 d\phi^2\nonumber \\
&=&- \frac{\ell^2}{(1+UV)^2} \left[ -4\;  dU dV + M \left(1-UV\right)^2 d\phi^2 \right]\label{SchwarzschildBTZ} 
\end{eqnarray}
respectively in Schwarzchild-like coordinates and Kruskal coordinates, where  ${t \in (-\infty,\infty)}$, ${r\in(0,\infty)}$,  ${\phi\in(0,2\pi)}$, and $r_h = \ell\sqrt{M}$   is the horizon. The null coordinates cover the entire maximally extended spacetime shown in figure \ref{fig:GeonConf}(a), where
 $-1<UV<1$. 

The associated geon of the BTZ black hole can be constructed by making appropriate identifications on the maximally extended BTZ spacetime \cite{Louko:1998hc,Louko:2004ej}
\begin{equation}
   J: \left(U, V, \phi \right) \rightarrow \left(V, U,  \phi+\pi )\right), \label{Jisometry} 
\end{equation}
which introduces a freely acting involutive isometry, whose group is $\Gamma := \left\{ {\rm Id}_{\rm BTZ}, J \right\} \simeq \mathbb{Z}_2$. The geon is the quotient space of the BTZ spacetime under this isometry:
it is isometric to one exterior region of the BTZ black hole,
as illustrated in figure~\ref{fig:GeonConf},
and so outside of the horizon, the geon spacetime is identical to the BTZ spacetime, sharing all of its local isometries. 
It is a spacetime with a topological twist analogous to a Klein bottle, with the twist hidden behind an event horizon.
The geon is time orientable, admitting a global foliation with spacelike hypersurfaces of topology $\mathbb{RP}^2 \backslash$ $\{$point at infinity$\}$, implying a similar change in the topology of the singularity. 
We choose the Hartle-Hawking vacuum, since it is invariant under the involution $J$, so it induces a unique vacuum on the geon background \cite{Louko:1998dj}.

The BTZ Wightman function for a conformally coupled quantum scalar field (in the Hartle Hawking vacuum) can be constructed from AdS$_3$ written in Rindler coordinates by identifying $\phi\to\phi + 2\pi n$ where $n$ is an integer. This yields the image sum over the AdS Wightman functions \cite{Lifschytz1994,Carlip:2003}
\begin{equation}
	W_{\tx{BTZ}}(x,x') = \frac{1}{4\pi\sqrt{2}\ell} \sum_{n=-\infty}^\infty \bigg( \frac{1}{\sqrt{\sigma_n}} - \frac{\zeta}{\sqrt{\sigma_n+2}}\bigg),
\end{equation}
where
\begin{multline}
	\sigma_n = \frac{rr'}{r_h^2} \cosh\bigg[ \frac{r_h}{\ell}(\Delta\phi-2\pi n) \bigg] - 1 \\
    -\frac{\sqrt{r^2-r_h^2} \sqrt{r'^2-r_h^2}}{r_h^2}\cosh\bigg[ \frac{r_h}{\ell^2}\Delta t -\iu\epsilon\bigg]
\label{eq:sigmabtz}
\end{multline}
with $\Delta\phi = \phi - \phi'$ and $\Delta t = t-t'$ and the square root is defined with the branch cut along the negative real axis \cite{Lifschytz1994}.
 The parameter $\zeta$ can take on values of $-1$, $0$, or $1$, corresponding to Neumann, transparent, or Dirichlet boundary conditions at spatial infinity respectively. 

The geon Wightman function can be found through a further image sum as \cite{Louko:1998dj}
\begin{multline}
	W_{\tx{Geon}}(x,x') = W_{\tx{BTZ}}(x,x') \\
    +\frac{1}{4\pi\sqrt{2}\ell} \sum_{n=-\infty}^\infty \bigg( \frac{1}{\sqrt{\tilde{\sigma}_n}} - \frac{\zeta}{\sqrt{\tilde{\sigma}_n+2}}\bigg),
\label{eq:wightman-geon}
\end{multline}
with
\begin{multline}
	\tilde{\sigma}_n = \frac{rr'}{r_h^2} \cosh\bigg[ \frac{r_h}{\ell}\big(\Delta\phi-2\pi (n+1/2)\big) \bigg] - 1 \\
    +\frac{\sqrt{r^2-r_h^2} \sqrt{r'^2-r_h^2}}{r_h^2}\cosh\bigg[ \frac{r_h}{\ell^2}(t+t') \bigg].
\label{eq:sigmageon}
\end{multline}

\section{Entanglement Harvesting for Detectors with Gaussian Switching Functions}
\label{sec:EHDGS}

We will now consider the entanglement harvesting protocol for two identical UDW detectors, with gaps $\Om_A=\Om_B=\Om$, that are located at fixed distances outside of the black hole
\eqn{
    x_D(\ta_D) \ce \Set{t=\frac{\ta_D}{\ga_D},\ r=R_D, \ph=\Ph_D}
}
where $\ga_D \ce \sqrt{R_D^2-r_h^2}/\ell$ is redshift factor of each detector $(D\in\{A,B\})$.  For simplicity of the parameter space, we will will also take $\Ph_A=\Ph_B=\Ph$.  Additionally, we will always label the detectors such that detector $A$ is the one closer to the horizon, $R_B>R_A$.  The proper separation between the two detectors is
\eqn{
    d(R_A,R_B) = \ell \log\lb(\frac{R_B+\sqrt{R_B^2+r_h^2}}{R_A+\sqrt{R_A^2+r_h^2}}\rb)
}
at a fixed $t$.

Finally, we define the switching function of each detector to be a Gaussian
\begin{equation}
	\chi(\tau)=e^{-\tau^2/2\sigma^2}.
\end{equation}

With the detectors' trajectories and switching functions determined, we are able to calculate the relevant detector quantities:
\aln{
    P_{D,\tx{Geon}} &= P_{D,\tx{BTZ}} + \De_{P_D} \label{eq:PDGeon} \\
    X_{\tx{Geon}} &= X_{\tx{BTZ}}+\De_{X}
    \label{eq:XGeon}
}
where, under the assumption that $\Ph_A=\Ph_B$,
\begin{widetext}
    \aln{
      P_{D,\text{BTZ}} &=  \frac{\lambda^2\sigma}{\sqrt{2\pi}} \vast\{\sqrt{\frac{\pi}{2}} \int_{-\infty}^{\infty} dy\ \frac{\ec^{-(\sigma\Omega-y)^2}}{\ec^{y/T}+1} - \frac{\zeta}{2} \Re\lb(\int_{0}^{\infty} dy\ \frac{\ec^{-a_Py^2}\ec^{-\iu\beta_Py}}{\sqrt{\cosh(\alpha_{P,0}^+)-\cosh(y)}}\rb) \nn\\
      & + \sum_{n=1}^{\infty}\vast[\Re\lb(\int_{0}^{\infty} dy\ \frac{\ec^{-a_Py^2}\ec^{-\iu\beta_Py}}{\sqrt{\cosh(\alpha_{P,n}^-) - \cosh(y)}}\rb) -\zeta \Re\lb(\int_{0}^{\infty} dy\ \frac{\ec^{-a_Py^2}\ec^{-\iu\beta_Py}}{\sqrt{\cosh(\alpha_{P,n}^+)-\cosh(y)}}\rb) \vast]\vast\} \label{eq:PDBTZ}\\
      \Delta_{P_D} &= \frac{\lambda^2\sigma^2}{4\sqrt{2\pi}} \ec^{-\sigma^2\Omega^2} \sum_{n=-\infty}^{\infty} \lb[\int_{-\infty}^{\infty} dy \frac{\ec^{-a_Py^2}}{\sqrt{\mathcal{Z}_{P,n}^-+\cosh(y)}} - \zeta \int_{-\infty}^{\infty} dy\ \frac{\ec^{-a_Py^2}}{\sqrt{\mathcal{Z}_{P,n}^++\cosh(y)}} \rb] \label{eq:DeltaPD} \\
      X_{\text{BTZ}} &= -\frac{\lambda^2\sigma}{2\sqrt{\pi}} \sqrt{\frac{\gamma_A\gamma_B}{\gamma_A^2+\gamma_B^2}}\exp\lb[-\sigma^2\Omega^2\lb(\frac{1}{2}+\frac{\gamma_A\gamma_B}{\gamma_A^2+\gamma_B^2}\rb)\rb] \nn\\
      &\quad\times \sum_{n=-\infty}^{\infty} \vast[\int_{0}^{\infty} dy\ \frac{\ec^{-a_Xy^2}\cos(\beta_X^{-}y)}{\sqrt{\cosh(\alpha_{X,n}^-)-\cosh(y)}} - \zeta \int_{0}^{\infty} dy\ \frac{\ec^{-a_Xy^2}\cos(\beta_X^{-}y)}{\sqrt{\cosh(\alpha_{X,n}^+)-\cosh(y)}}\vast] \label{eq:XBTZ}\\
      \Delta_X &= -\frac{\lambda^2\sigma}{2\sqrt{\pi}}\sqrt{\frac{\gamma_A\gamma_B}{\gamma_A^2+\gamma_B^2}}\exp\lb[-\sigma^2\Omega^2\lb(\frac{1}{2}-\frac{\gamma_A\gamma_B}{\gamma_A^2+\gamma_B^2}\rb)\rb] \nn\\
      &\qquad\qquad \times \sum_{n=-\infty}^{\infty}\lb[\int_{0}^{\infty} dy\ \frac{\ec^{-a_X y^2}\cos(\beta_X^{+}y)}{\sqrt{\mathcal{Z}_{X,n}^-+\cosh(y)}} - \zeta \int_{0}^{\infty} dy\ \frac{\ec^{-a_Xy^2}\cos(\beta_X^{+}y)}{\sqrt{\mathcal{Z}_{X,n}^++\cosh(y)}}\rb] \label{eq:DeltaXNoAngle}
}
\end{widetext}
and
\aln{
    T &\ce \frac{r_h\sigma}{2\pi\ell\sqrt{R_D^2-r_h^2}} \nn\\
    a_P &\ce \frac{\ga_D^2\ell^4}{4\si^2r_h^2} \nn\\
    \be_P &\ce \frac{\ga_D\Om\ell^2}{r_h} \nn\\
    \al_{P,n}^{\pm} &\ce \arccosh\lb[\frac{r_h^2}{\ga_D^2\ell^2}\lb(\frac{R_D^2}{r_h^2}\cosh\lb(\frac{r_h}{\ell}2\pi n\rb) \pm 1\rb)\rb] \nn\\
    \mc{Z}^\pm_{P,n} &\ce \frac{r_h^2}{\ga_D^2\ell^2}\lb(\frac{R_D^2}{r_h^2}\cosh\lb(\frac{r_h}{\ell}2\pi(n+1/2)\rb)\pm 1\rb)
}
\aln{
    a_X &\ce \frac{1}{2\si^2}\frac{\ga_A^2\ga_B^2}{\ga_A^2+\ga_B^2}\frac{\ell^4}{r_h^2} \nn\\
    \be_X^\pm &\ce \frac{\Om\ga_A\ga_B(\ga_A\pm\ga_B)}{\ga_A^2+\ga_B^2}\frac{\ell^2}{r_h} \nn\\
    \al_{X,n}^{\pm} &\ce \arccosh\lb[\frac{r_h^2}{\ga_A\ga_B\ell^2}\lb(\frac{R_AR_B}{r_h^2}\cosh\lb(\frac{r_h}{\ell}2\pi n\rb) \pm 1\rb)\rb] \nn\\
    \mc{Z}^\pm_{X,n} &\ce \frac{r_h^2}{\ga_A\ga_B\ell^2}\lb(\frac{R_AR_B}{r_h^2}\cosh\lb(\frac{r_h}{\ell}2\pi(n+1/2)\rb)\pm 1\rb).
}
We evaluate these integrals numerically,  using the \texttt{DoubleExoponetial} method of Mathematica to a working precision and accuracy of 20. We will consider Dirichlet boundary conditions, $\zeta = 1$, for the remainder of our discussion, since all three boundary conditions produce qualitatively similar results. Finally, we define the unit less coupling constant $\ti{\la}\ce\la\sqrt{\si}$ for convenience.

\subsection{Transition Probability in Geon Spacetime}

\begin{figure*}[ht]
\includegraphics[width=0.33\textwidth]{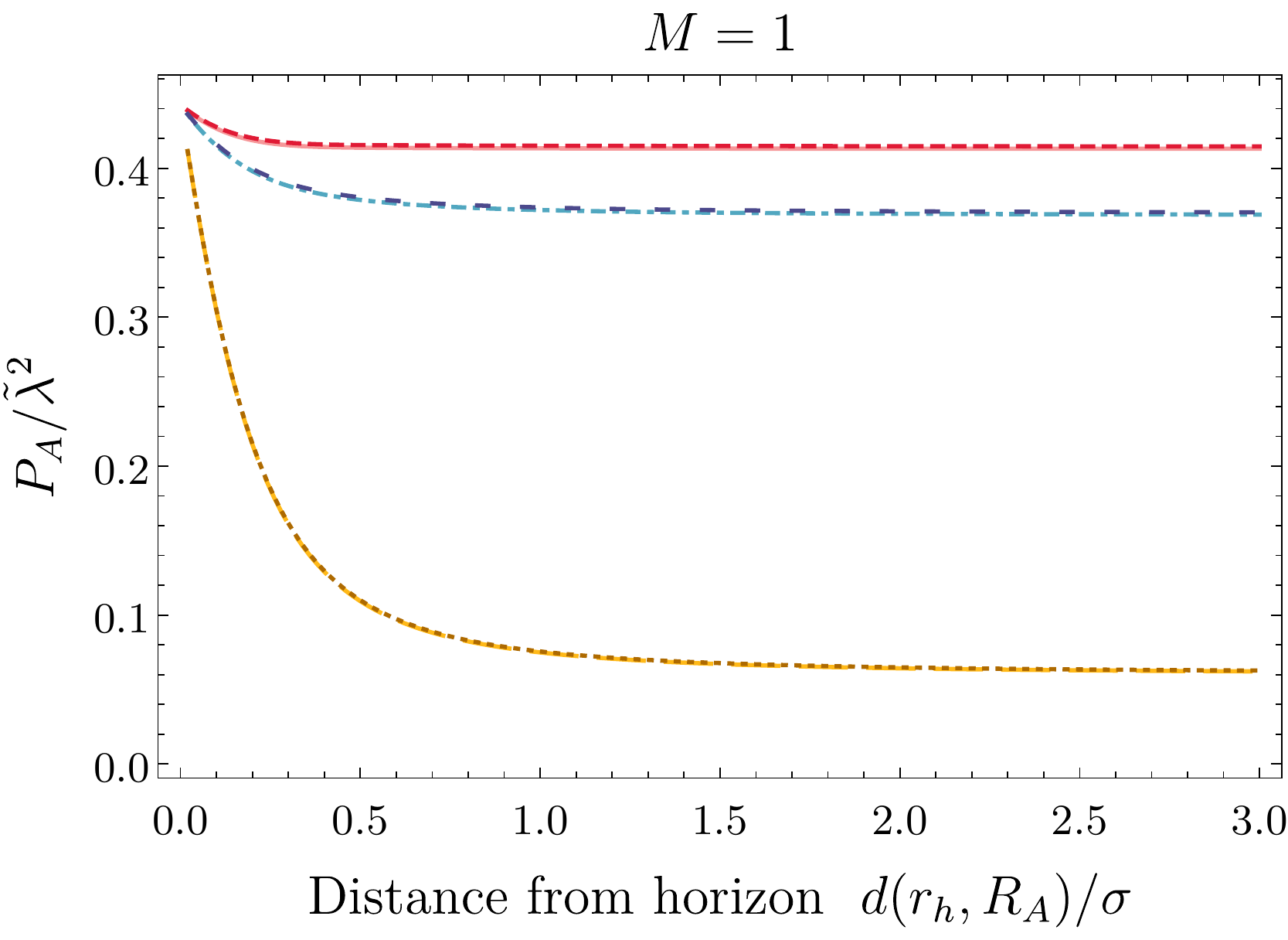}
\hspace{1em}
\includegraphics[width=0.33\textwidth]{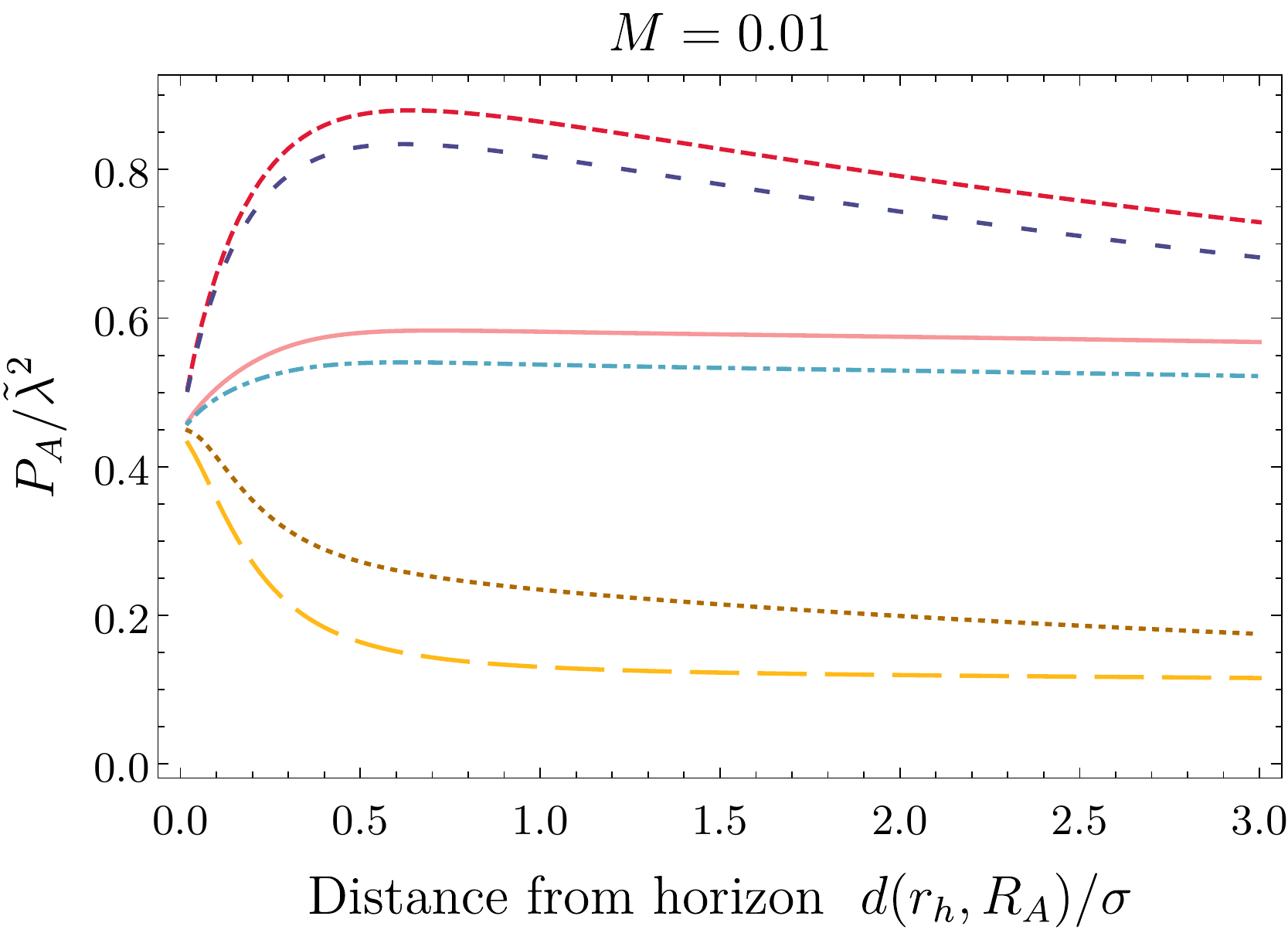}
\hspace{1em}
\includegraphics[width=0.15\textwidth]{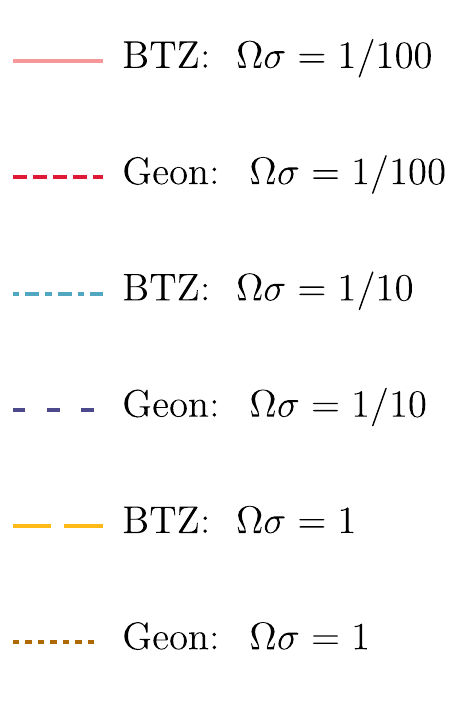}
\caption{%
    A comparison of the transition probability as a function of the detector's proper distance from the horizon in a BTZ and geon spacetime for various values of the energy gap with a black hole mass of \textit{(left)} $M=1$ and \textit{(right)} $M=0.01$.  When the black hole mass is large, the transition probability is almost exactly the same in both spacetimes, but when the mass is small, the transition probability is larger in a geon spacetime than a BTZ spacetime.  The AdS length is set to $\ell=10\si$.
}
\label{fig:PDvsPropRD}
\end{figure*}

First we consider the dependence of the transition probability of a single UDW detector on its proper distance from the horizon in BTZ and geon spacetimes, which is shown in figure \ref{fig:PDvsPropRD}.  We find that when the black hole mass is large, $M=1$, the transition probabilities of the detector are nearly identical in both spacetimes.  However, when the mass is low, $M=0.01$, we find that the transition probability is significantly larger in a geon spacetime than in the  BTZ spacetime, regardless of the energy gap of the detector.  We also find that as the detector approaches the horizon, the transition probability approaches the same value $(P_D/\ti{\la}^2 \approx 0.45)$ independent of the spacetime, the black hole mass, or the detector's energy gap.

\begin{figure}
\includegraphics[width=\linewidth]{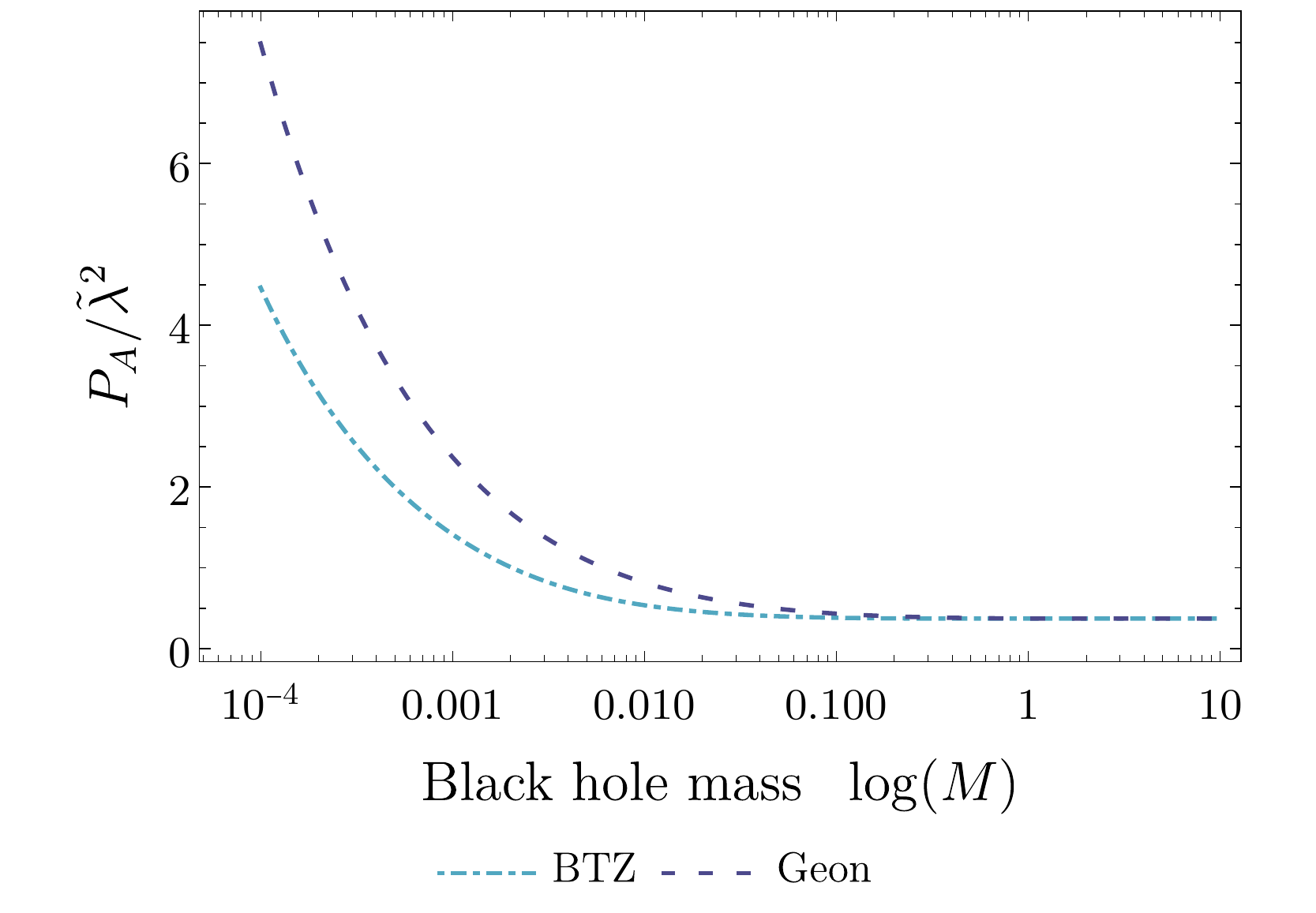}
\caption{%
    A comparison of the transition probability as a function of the black hole mass in a BTZ and geon spacetime for a black hole mass.  When the black hole mass is large, the transition probability is nearly identical in both spacetimes. The transition probability increases as the black hole mass decreases and increases faster in a geon spacetime.  The detector has an energy gap of $\Omega\sigma=0.1$ and is fixed at a proper distance of $d(r_h,R_A)=\sigma$ from the horizon and the AdS length is set to $\ell=10\si$.
}
\label{fig:PDvsLogM}
\end{figure}

The transition probability can only be used to distinguish between the two spacetimes in the low mass regime due to the properties of the image sum (Eqns.\ \eqref{eq:PDBTZ} and \eqref{eq:DeltaPD}).  For large values of $n$, the integrands are dominated by $\ec^{-\pi\sqrt{M}n}$, so are exponentially suppressed in mass.  It turns out that when $M=1$, the image sum is dominated by the $n=0$ term, so the spacetime is essentially AdS-Rindler, and any black hole effects are negligible \cite{AH:plb2020}.  However, when the  mass is small enough, the higher order terms begin to contribute significantly to the Wightman function, and will have an effect on the final state of the detector.  This is highlighted in figure \ref{fig:PDvsLogM}, where we plot the transition probability versus the black hole mass, and find that the smaller the mass, the larger the difference between the transition probability in geon spacetime and  BTZ spacetime.

\subsection{Concurrence in Geon Spacetime}

\begin{figure}[th]
\includegraphics[width=\linewidth]{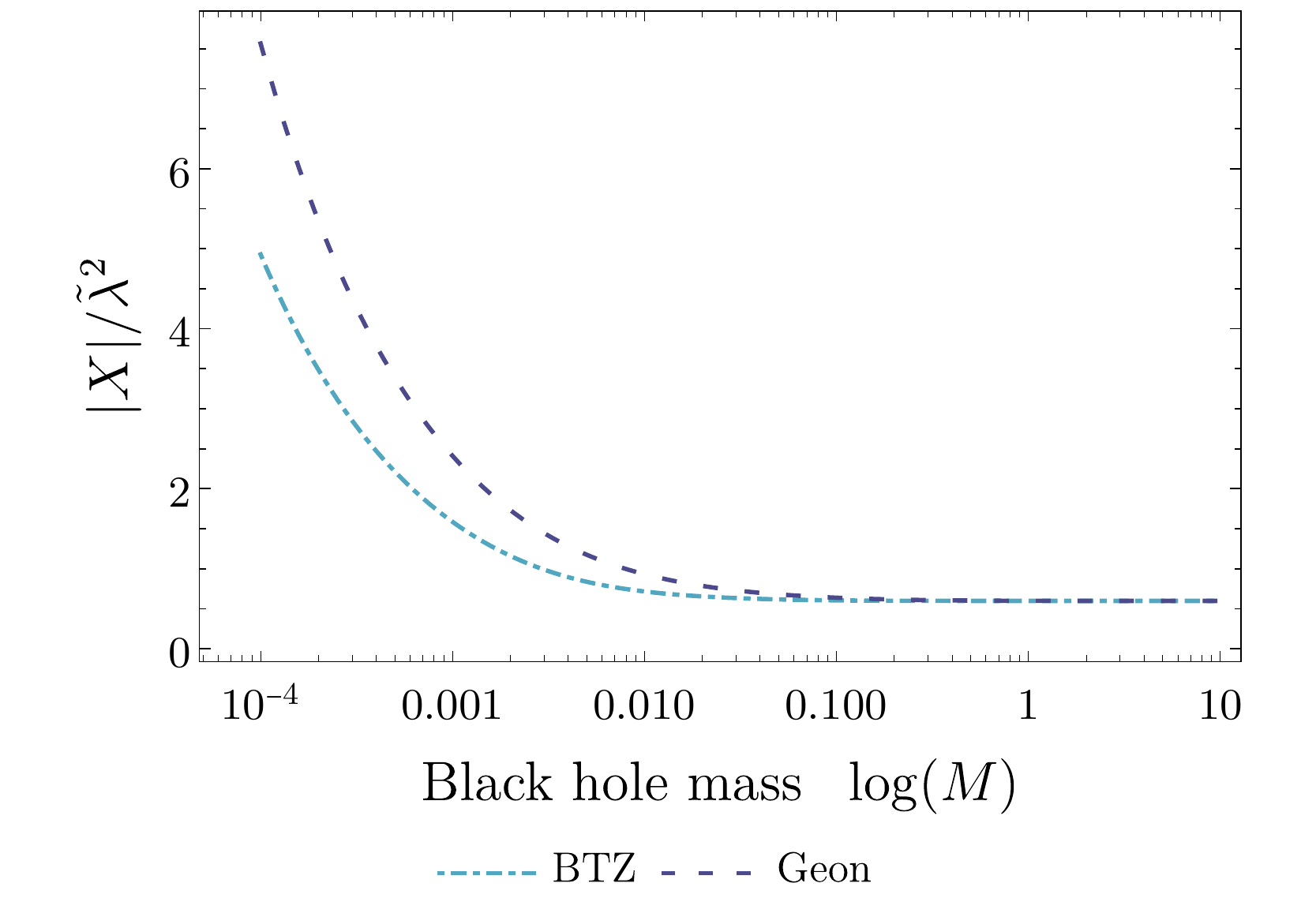}
\caption{%
     A comparison of the non-local correlations, described by the matrix element $|X|$, as a function of the black hole mass in a BTZ and geon spacetime for a black hole mass.  When the black hole mass is large, the non-local correlations are nearly identical in both spacetimes. The value of $|X|$ increases as the black hole mass decreases and increases faster in a geon spacetime.  The detectors have an energy gap of $\Omega\sigma=0.1$, are separated by a proper distance $S=0.5\si$, and detector $A$ is located at a proper distance of $d(r_h,R_A)=\sigma$ from the horizon and the AdS length is set to $\ell=10\si$.
}
\label{fig:XvsLogM}
\end{figure}

Now we consider the entanglement harvested by a pair of identical UDW detectors in the geon spacetime as compared to the BTZ spacetime. Recall, the entanglement of the two detectors depends on the transition probabilities of the two detectors and the matrix element $X$, which encodes the non-local correlations between them.

  In figure \ref{fig:XvsLogM} we plot the dependence of the non-local correlations, $X$, on the black hole mass in the BTZ and   geon spacetimes.  Similar to the transition probability, we find that when the mass is large, the non-local correlations of the two detectors in a geon spacetime are almost indistinguishable from the BTZ spacetime.  In both spacetimes, the non-local correlations increase as the black hole mass decreases, with the correlations incresesing faster in a geon spacetime.  Therefore, we expect that any differences in the entanglement harvested by the UDW detectors will occur in the small mass regime.

\begin{figure}[t]
  \centering
  \includegraphics[width=\linewidth]{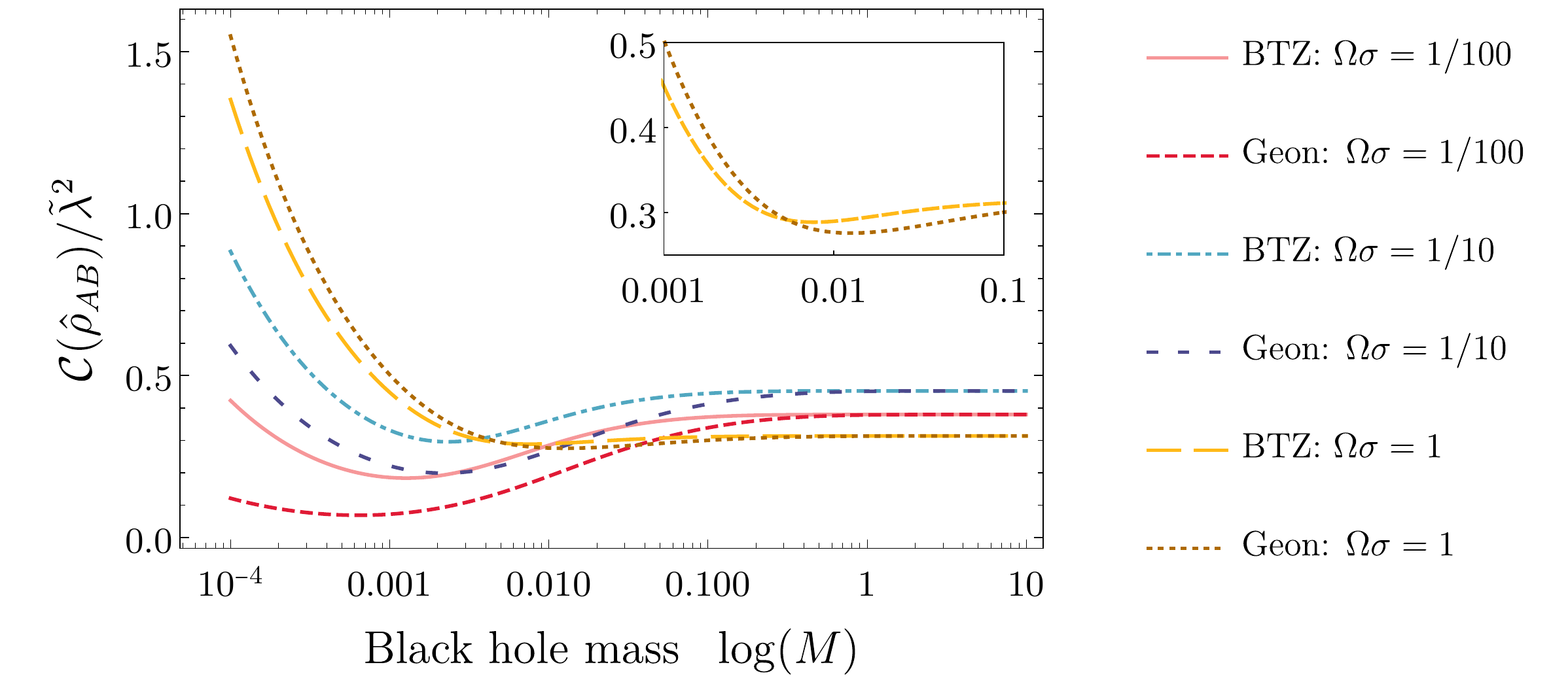}
  \caption{%
    The concurrence, $\mathcal{C}(\hat{\rho}_{AB})/\lambda^2$, of the density matrix as a function of the black hole mass for various values of the energy gap.  When the mass is large (close to or greater than 1), the Wightman function of the fields is dominated by the $n=0$ term in the image sum, and the detectors behave as if they were in AdS-Rindler space.  Detectors with a small energy gap can harvest more entanglement in a black hole spacetime and detectors with a large energy gap harvest more entanglement in a geon spacetime.  The inset shows detail for $\Om\si=1$.  Detector $A$ is a proper distance of $d(r_h,R_A)=\si$ from the horizon, the proper separation of the detectors is $S=0.5\si$, and the AdS length is set to $\ell=10\si.$
  }
  \label{fig:CvsLogM}
\end{figure}

We now consider the effect of mass on  the entanglement harvested from both the BTZ and geon spacetimes, which we plot in figure \ref{fig:CvsLogM}. As we saw with the transition probability and the non-local corrlations, the two spacetimes can only be distinguished through the entanglement harvesting protocol in the low mass regime.  However, the difference in the entanglement is much less dramatic than was seen for the individual matrix elements.  By comparing figures \ref{fig:PDvsLogM} and \ref{fig:XvsLogM}, it can been seen that $P_D$ and $|X|$ in both the BTZ and geon spacetimes increase at very similar rates with decreasing mass, and they are both higher in the geon spacetime.  As a result, the amount of entanglement harvested in both spacetimes will be similar, though not identical in the case of small black hole mass.  The detectors are more correlated (in terms of $|X|$), but are also noisier in a geon spacetime.

The ``best'' spacetime for harvesting entanglement depends on the energy gap of the detector.  When the energy gap is small $(\Om\si=0.01)$, more entanglement can be harvested in a BTZ spacetime than a geon spacetime for the same detector configuration.  However, when the energy gap is large $(\Om\si=1)$ the ``best spacetime'' is also mass dependent; more entanglement can be harvested in a geon spacetime when the mass is very small $(M\lessapprox0.002)$, and more entanglement can be harvested in a BTZ spacetime when the mass is moderately small.

In both spacetimes, and for the explored values of $\Om$,  we find that  as the mass of the black hole decreases, the amount of entanglement the detectors are able to harvest from the vacuum decrease to some minimum value.  The position of the minimum is spacetime and energy gap dependent.  As the mass is decreased further, the detectors are able to extract more entanglement from the field.

\begin{figure}[t]
  \centering
  \includegraphics[width=\linewidth]{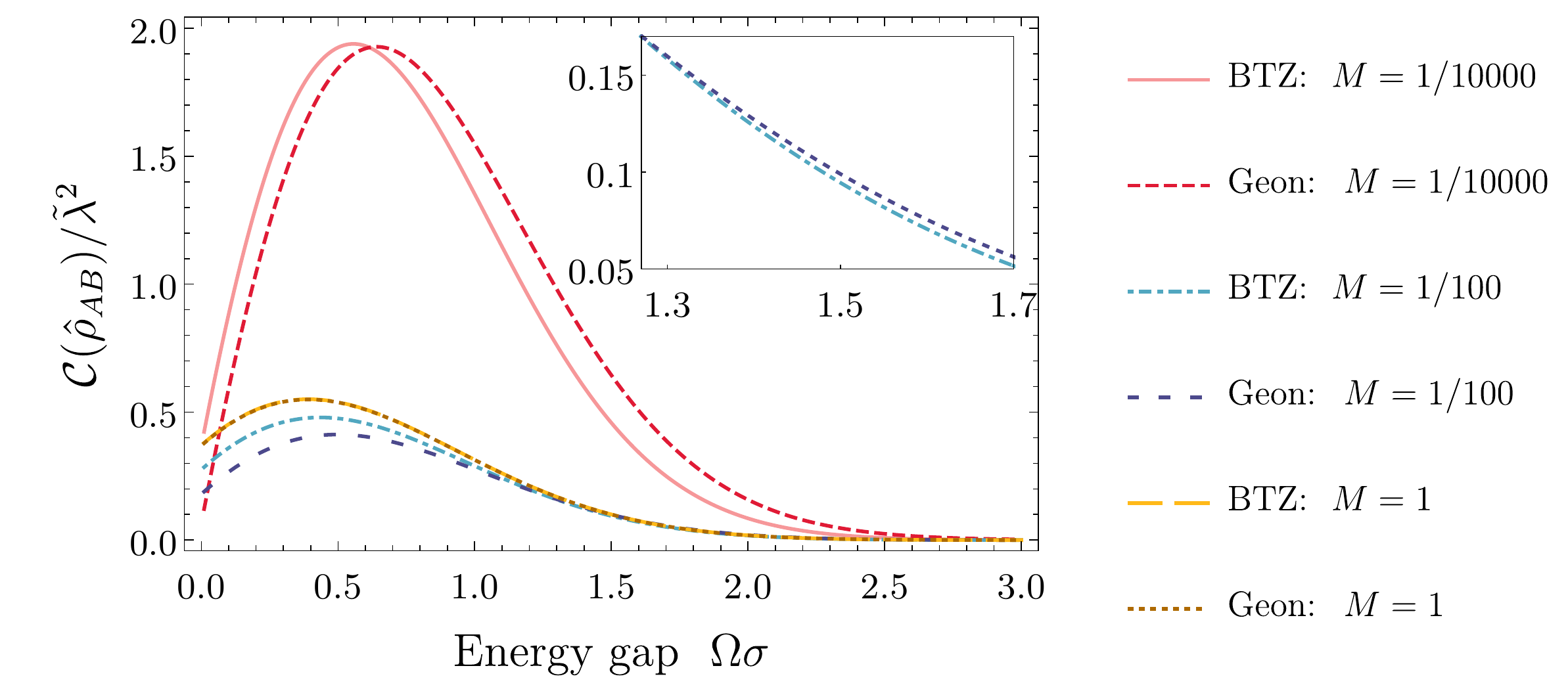}
  \caption{%
    The concurrence of the reduced density matrix, $\mathcal{C}(\hat{\rho}_{AB})/\lambda^2$, as a function of the energy gap of the detector for various blackhole and geon masses.  Detectors with a small energy gap can harvest more entanglement in a black hole spacetime and detectors with a large energy gap harvest more entanglement in a geon spacetime.  The energy gap where this crossover happens depends on the mass.  The inset shows the crossover for $M=0.01$. The proper distance of detector $A$ from the horizon is $d(r_h,R_A)=\sigma$, the proper distance between detector $A$ and $B$ is $d(R_A,R_B)=0.5\sigma$, and the AdS length is $\ell=10\sigma$. %
  }
  \label{fig:CvsOmega}
\end{figure}

We further explore the dependence of the ``best'' spacetime for entanglement harvesting on the energy gap of the detector in figure \ref{fig:CvsOmega}.  In general, we find that when the energy gap of the detectors is small, they are able to harvest more entanglement in a BTZ spacetime than in its geon counterpart.  However, the when the energy gap of the detector is large, more entanglement can be harvested in a geon spacetime.  The actual energy gap of where this switch occurs is dependent on the mass of the black hole.

Mathematically, we can explain much of this behaviour by studying the prefactors of the expressions for $|X|$ and $P_D$.  First, looking at Eq.\ \eqref{eq:XBTZ}, we see that the non-local correlations in a BTZ spacetime are exponentially suppressed with increasing values of the energy gap.  However, in Eq.\ \eqref{eq:DeltaXNoAngle}, the redshift factors contribute positively, rather than negatively, in the exponential, meaning that the energy gap will have less of an effect on the non-local correlations in a geon spacetime.  This change of sign is a direct result of the geon definitions taking $U\leftrightarrow V$.  It is important to note that the relativistic quantity
\eqn{
    \frac{\ga_A\ga_B}{\ga_A^2+\ga_B^2} \le \frac{1}{2}
    \label{eq:RSF}
}
with saturation occurring when $\ga_A=\ga_B$, so the overall non-local correlations in a geon spacetime can never grow without bound with increasing energy gap.  However, we estimate that the larger the value of the energy, the more the geon correction will dominate the non-local correlations. 

Upon considering the transition probability we find that the energy gap has the opposite effect.  There is an overall $e^{-\si^2\Om^2}$ prefactor in front of the geon corrections (Eq.\ \eqref{eq:DeltaPD}) to the transition probability that is not present in expression for the transition probability in a BTZ spacetime (Eq.\ \eqref{eq:PDBTZ}).  From this we can estimate that as the energy gap of the detectors increases, the relative difference in the local noise in the two spacetimes will decrease.  Overall we find that for large energy gaps, the non-local correlations between the detectors is much larger in a geon spacetime, but the local noise will be comparable between the two spacetimes. Therefore detectors with large enough energy gaps should be able to harvest more entanglement in the geon spacetime.  However, these prefactors do not capture all of the energy gap dependence since there is sill dependence within the integrands themselves.

 The peak in concurrence that occurs for both large and small black hole mass is a generic property of the entanglement harvesting protocol. The exact position of the peak depends on the configuration of the detectors and the underlying spacetime, but the physical the intuition is as follows: when the energy gaps of the detectors are small, they are very sensitive to local fluctuations of the field, leading to a large transition probabilities and little to no entanglement harvested.  Increasing the energy gap(s) will decrease the local noise of the detectors; however, this will also decrease their sensitivity  to correlated field fluctuations, leading to a decrease in the non-local correlations, $|X|$.  Since the transition probability and the non-local correlations decrease at different rates as the energy gaps are increased, it is possible to tune the energy gaps of the detectors to maximise concurrence.

\begin{figure*}[t]
  \centering
  \includegraphics[width=0.33\textwidth]{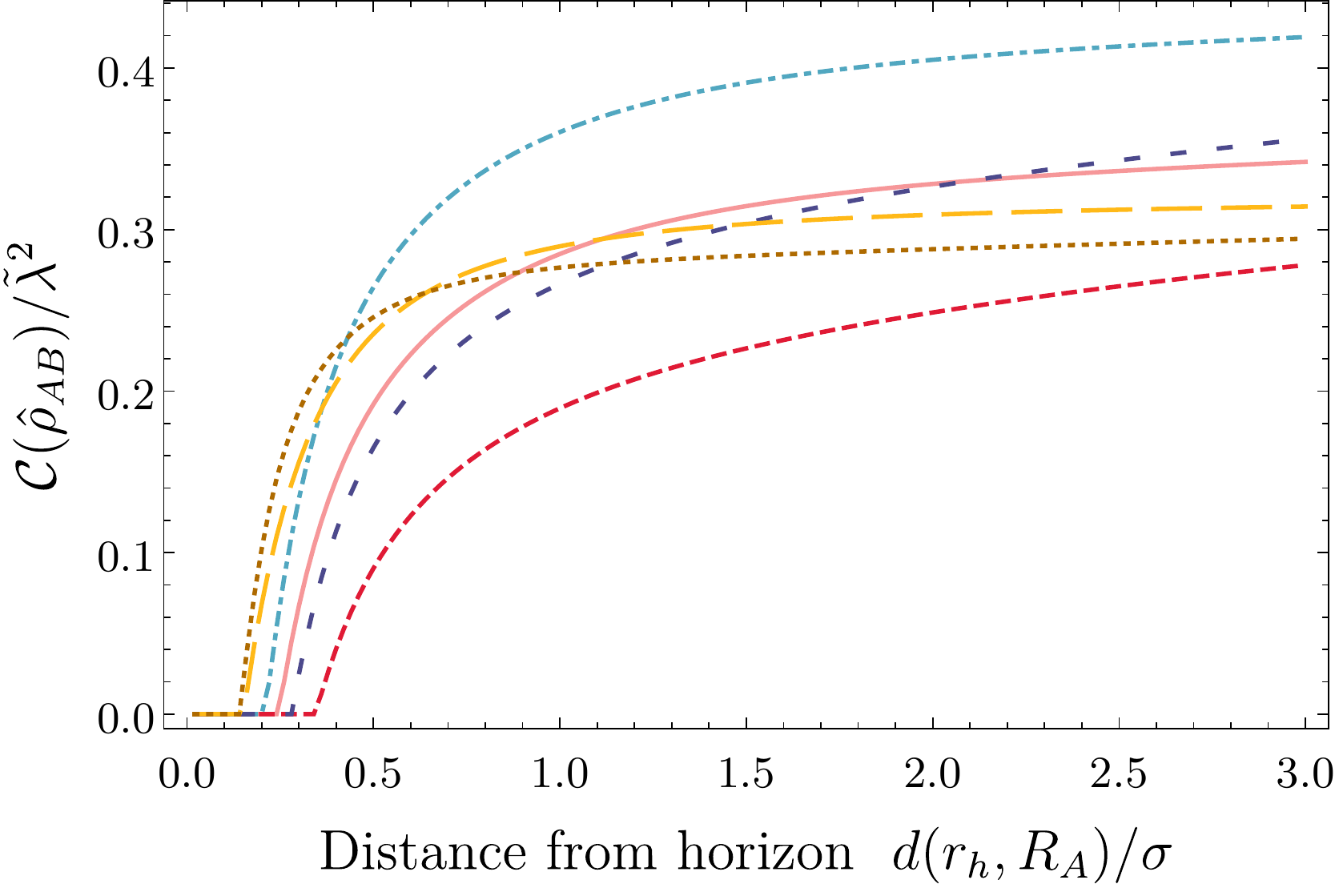}
  \hspace{1em}
  \includegraphics[width=0.33\textwidth]{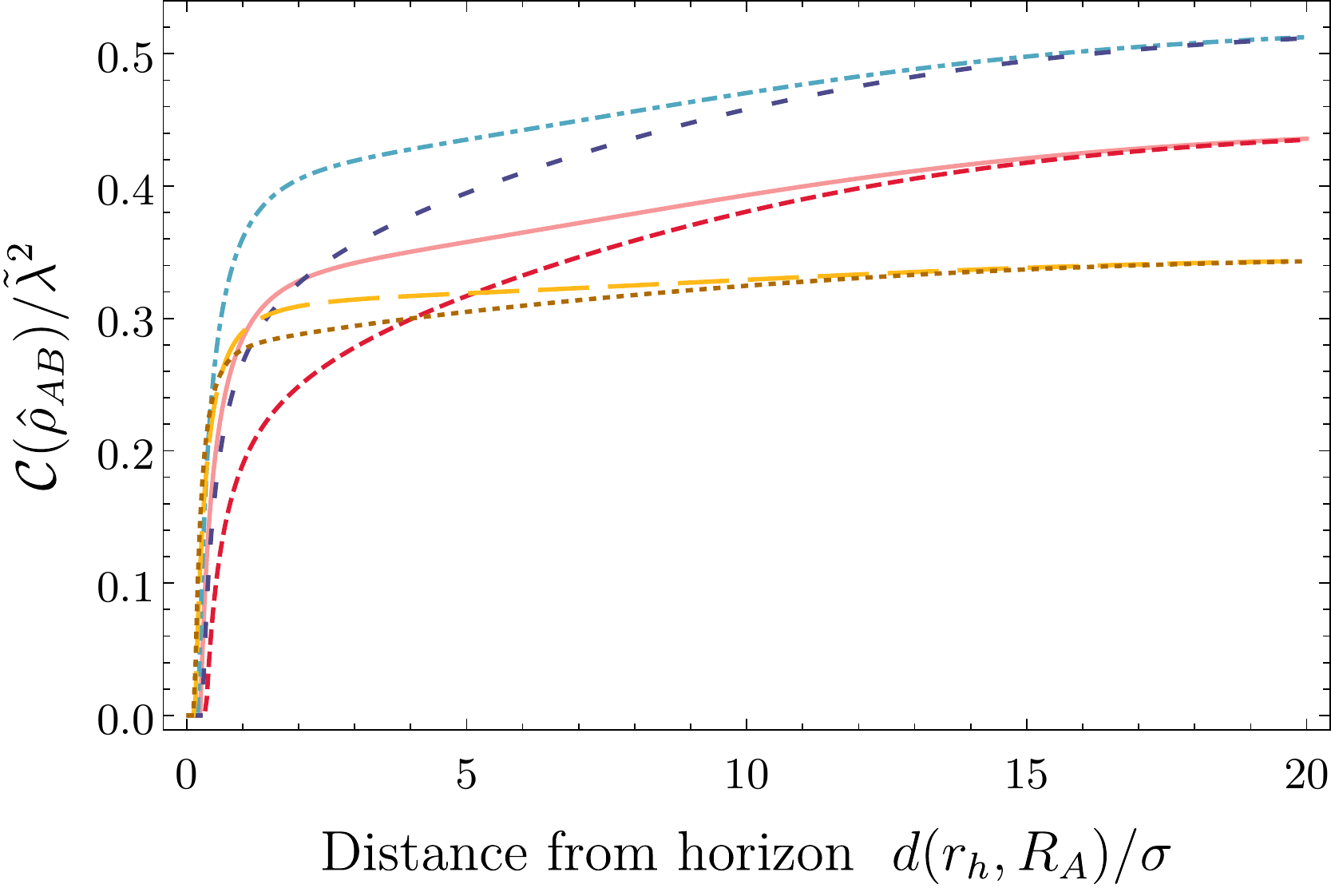}
  \hspace{1em}
  \includegraphics[width=0.15\textwidth]{Legend_OmegaV.pdf}
  \caption{%
    The concurrence of the density matrix, $\mathcal{C}(\hat{\rho}_{AB})/\lambda^2$, as a function of the proper distance of detector $A$ from the horizon of the detector, $d(r_h,R_A)$, for various detector energy gaps in a BTZ and a geon spacetime.  The left plot shows the behaviour close to the horizon, and the right plot shows a larger range of distances.  The proper separation of the detectors is set to $S=0.5\si$, the black hole mass is $M=0.01$, and the AdS length is $\ell=10\si$.
  }
  \label{fig:CvsPropRA}
\end{figure*}

In figure \ref{fig:CvsPropRA}, we plot the dependence of the resulting entanglement of the two detectors on the proper distance of detector $A$ from the horizon while keeping the proper separation of the detectors fixed.  We find that for all detector energy gaps, the detectors harvest more entanglement in a BTZ spacetime than a geon spacetime when they are somewhat close $d(r_h,R_A)\gtrapprox0.5\si$ to the horizon.  When the detectors are far away from the horizon, the resulting entanglement of the detectors is approximately the same in both spacetimes.  This makes intuitive sense, since the detectors are far away from the topological difference in the two spacetimes where the impact on the field is expected to be reduced.

However, when the energy gap is large, $\Om\si=1$, and the detectors are very close to the horizon, they are able to harvest more entanglement in the geon spacetime. Such a crossover never occurs for detectors with a smaller energy gap.

The reason for the large energy gap behaviour can be understood by again looking at the prefactors of Eqs.~\eqref{eq:XBTZ} and~\eqref{eq:DeltaXNoAngle}.  The positive contribution of the exponential in the geon contribution to the non-local correlations is dependent on the relative redshift between the two detectors; the smaller the relative redshift, the larger this positive contribution will be.  When the detectors are placed further from the horizon, while keeping the proper separation between them fixed, the relative redshift of the detectors decreases.  As a result, the geon contribution to the non-local correlations becomes larger, most significantly for   large values of the energy gap.  Therefore, in this part of the parameter space, detectors are able to harvest more entanglement in a geon spacetime.

Since detectors are able to harvest more entanglement in a BTZ spzacetime when they are close to the horizon, we also find a difference in the size of the entanglement shadow\cite{henderson:2018,Robbins:2020jca}, the region near the horizon where detectors of with a specified proper separation and energy gap are not able to become entangled through the entanglement harvesting protocol.  For detectors with a small energy gap, the entanglement shadow is larger in a geon spacetime, which is consistent with them being able to harvest less entanglement in that spacetime.

\begin{figure*}[t]
  \centering
  \includegraphics[width=0.48\linewidth]{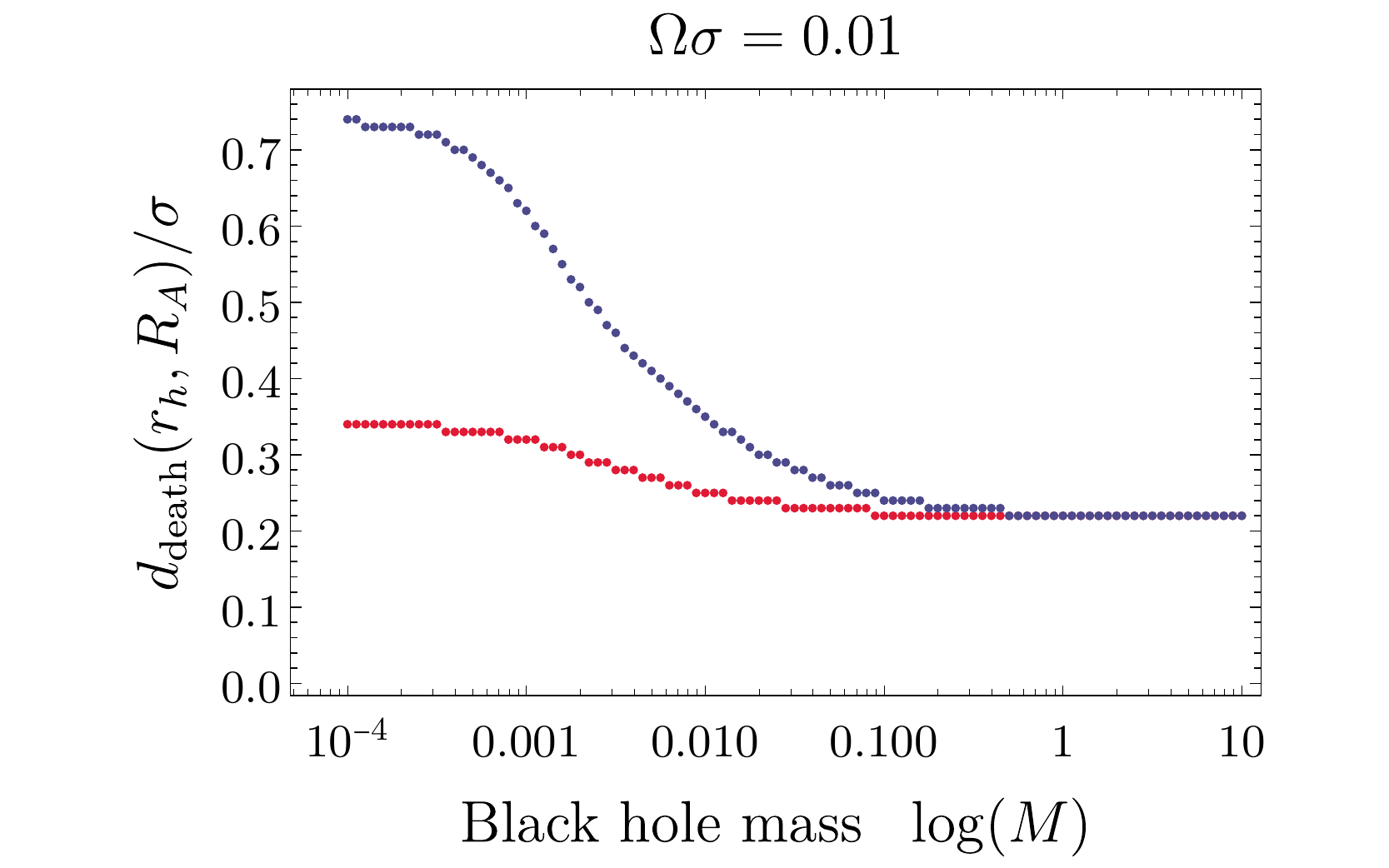}
  \includegraphics[width=0.48\linewidth]{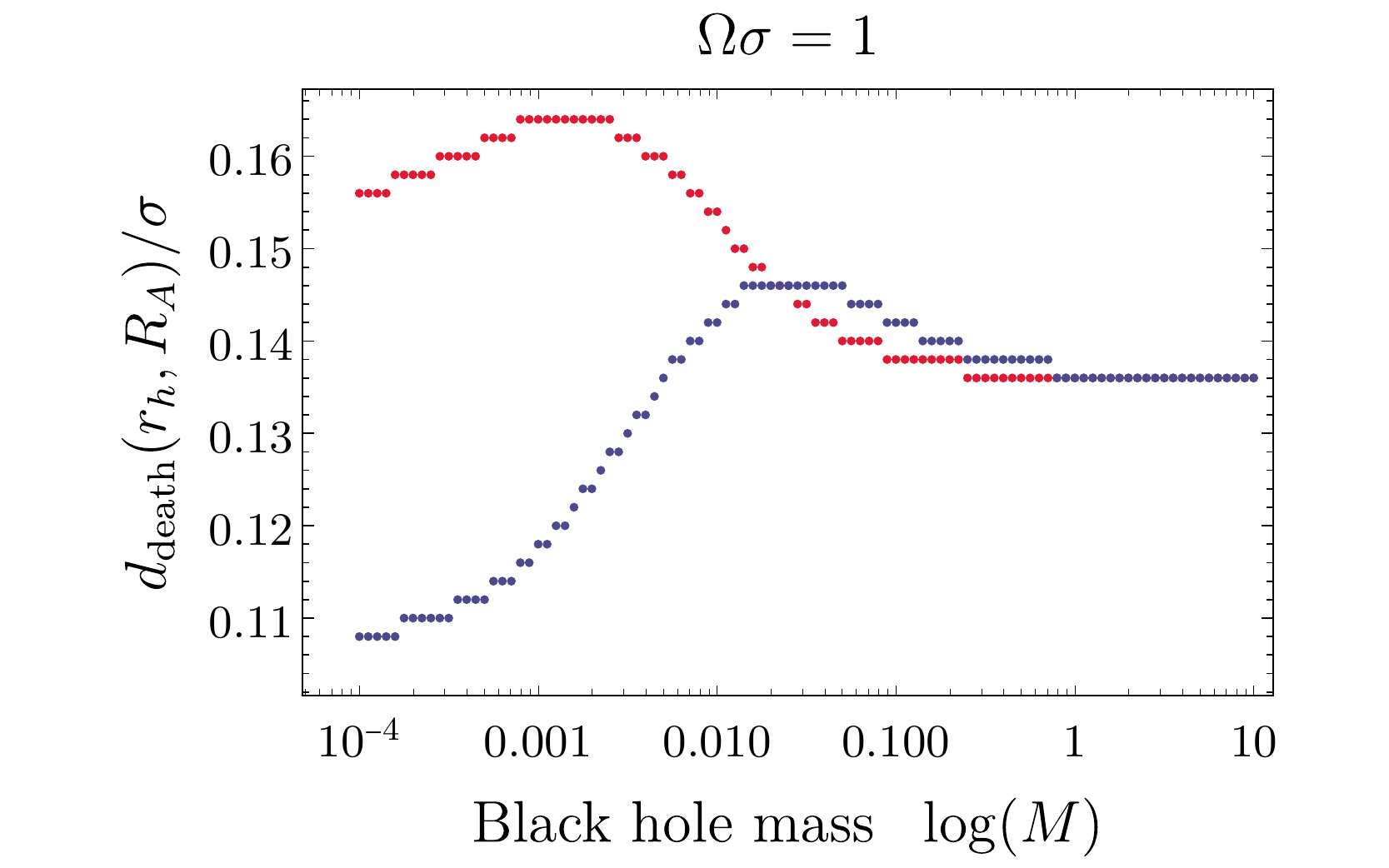}\\
  \vspace{0.5em}
  \includegraphics[width=0.55\linewidth]{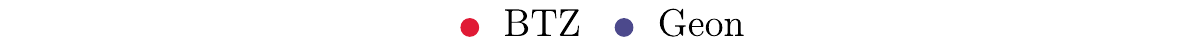}
  \caption{%
    The proper distance of detector $A$ from the horizon corresponding to the point where the detectors can no longer harvest entanglement, $d_{\tx{death}}(r_h,R_A)$ as a function of the black hole mass for an energy gap \textit{(left)} $\Om\si=0.01$ and \textit{(right)} $\Om\si=1$.  When the mass is large, $d_{\tx{death}}$ is the same in both the BTZ and geon spacetimes.  For small energy gaps, $d_{\tx{death}}$ is larger in the geon spacetime, corresponding to a larger entanglement shadow.  For  large energy gaps, $d_{\tx{death}}$ is smaller in the geon spacetime for very small mass black holes, but larger for moderate mass.  The proper separation of the detectors is set to $S=0.5\si$ and the AdS length is $\ell=10\si$.
  }
  \label{fig:ddeathvsLogM}
\end{figure*}

This impact is explored further in figure \ref{fig:ddeathvsLogM}, where we plot $d_\tx{death}(r_h,R_A)$, the proper distance of detector $A$ from the horizon corresponding to the point where the detectors are no longer able to become entangled, as a function of black hole mass; this is the boundary of the entanglement shadow.
When the energy gap  is small, $\Om\si=0.01$, the entanglement shadow is larger, corresponding to a larger value of $d_\tx{death}(r_h,R_A)$, in a geon spacetime.  However when the detector energy gap is large, $\Om\si=1$, the spacetime with the larger entanglement shadow is mass dependent.  It is larger in a BTZ spacetime for small mass blackholes $(M\lessapprox0.01)$, but larger in geon spacetime for moderate mass black holes.  When the mass of the blackhole is large $M\gtrapprox1$, the entanglement shadow is the same for both spacetimes, since in both cases, the spacetime is dominated by AdS-Rindler.

When the detectors are very close to the horizon, the large relative redshift between the detectors leads to a reduction of both the BTZ non-local correlations but also the geon correction. However from figure \ref{fig:PDvsPropRD}, the transition probabilities of the detectors remains high, being  much higher in the geon spacetime.  It therefore  makes sense that the geon has  a larger entanglement shadow.  

\section{Conclusion}
\label{sec:concl}

Consistent with Sir Roger's intuition, topology can have a significant effect on  physics.  We have seen in our investigation that topological structures hidden behind event horizons that are classically undetectable by outside observers leave detectable imprints on
the spacetime entangling properties of quantum fields.

We demonstrated this by comparing  the resulting entanglement of a pair of identical UDW detectors following a time dependent interaction with a conformally coupled massless scalar field in the Hartle-Hawking vacuum on a BTZ spacetime and its associated geon.  We found that if the mass of the black hole is large, then there is no noticeable difference in the entanglement extracted by the two detectors, or of the transition probability of a single detector, in either spacetime. 

When the mass of the black hole is small, we find that the transition probability of a single UDW detector is higher in a geon spacetime than a BTZ spacetime.  Furthermore, the smaller the black hole mass, the larger the difference.  Overall, this pattern holds when the distance from the detector to the horizon is varied; the transition probability in a geon spacetime remains larger.  One consequence is a more dramatic reduction  in the geon spacetime of the transition probability as the detector is placed closer to the   horizon, where the local temperature is higher.  This suggest that there the anti-Hawking effect may be found over a larger range of parameters in the geon spacetime \cite{AH:plb2020}.  We leave this exploration   for future work.

Although the non-local correlations between the two detectors are also lower in  a BTZ spacetime as compared to its geon counterpart,
this does not guarantee that the detectors harvest more entanglement in the latter setting.  In fact, since the dependence of the non-local correlations on mass is nearly identical to the local noise (transition probability) of the detectors, the relationship between the resulting entanglement of the detector to the mass is not   straightforward.

Detectors with a small energy gap harvest less entanglement in a geon spacetime as compared to the BTZ spacetime.  But when the energy gap is large, the detector is able to harvest more entanglement in a geon spacetime provided the mass is low enough.  In fact, for a given value of (small) mass, there is a value of the energy gap such that detectors harvest more mass in a geon spacetime.  One reason for this is that the geon corrections to the transition probability decrease overall with increasing energy gap, but  geon corrections to the non-local correlations increase with gap relative to those in the BTZ case.  We can conjecture that a vacuum quantum field in the geon spacetime has less local fluctuations but more non-local correlations   for
high frequency field modes,  and more local fluctuations and fewer non-local correlations for low frequency modes as compared to the BTZ spacetime.

Since detectors with low energy gap harvest less entanglement in a geon spacetime, we also find that this corresponds to a larger entanglement shadow around a geon.  Conversely, we find that   detectors with   large energy gap experience a smaller entanglement shadow  around the geon, provided the mass is small enough.  This difference is almost entirely due to the relative reduction in the local noise of the detectors in the geon spacetime.

Throughout our study we have only considered a pair of detectors with no relative angle between them.  The methods here can be easily extended to the case where $\De\ph\ne0$.  In a BTZ spactime it has been shown that the amount of entanglement harvested by a pair of detectors depends on the angle between them, and the relative angle which maximizes entanglement harvesting depends on the energy gap \cite{henderson:2021}.  Since the optimal spacetime for entanglement harvesting also depends on the energy gap, it would be interesting to study how the optimal angle between detectors changes in a geon spacetime.  We are likely to find further imprints of hidden topology here!

\section*{Acknowledgements}
This work was supported in part by the
Natural Science and Engineering Research Council of Canada (NSERC) and by Asian Office of Aerospace Research and Development Grant No.\ FA2386-19-1-4077.

\bibliography{LongPaper} %

\begin{thebibliography}{62}%
\makeatletter
\providecommand \@ifxundefined [1]{%
 \@ifx{#1\undefined}
}%
\providecommand \@ifnum [1]{%
 \ifnum #1\expandafter \@firstoftwo
 \else \expandafter \@secondoftwo
 \fi
}%
\providecommand \@ifx [1]{%
 \ifx #1\expandafter \@firstoftwo
 \else \expandafter \@secondoftwo
 \fi
}%
\providecommand \natexlab [1]{#1}%
\providecommand \enquote  [1]{``#1''}%
\providecommand \bibnamefont  [1]{#1}%
\providecommand \bibfnamefont [1]{#1}%
\providecommand \citenamefont [1]{#1}%
\providecommand \href@noop [0]{\@secondoftwo}%
\providecommand \href [0]{\begingroup \@sanitize@url \@href}%
\providecommand \@href[1]{\@@startlink{#1}\@@href}%
\providecommand \@@href[1]{\endgroup#1\@@endlink}%
\providecommand \@sanitize@url [0]{\catcode `\\12\catcode `\$12\catcode
  `\&12\catcode `\#12\catcode `\^12\catcode `\_12\catcode `\%12\relax}%
\providecommand \@@startlink[1]{}%
\providecommand \@@endlink[0]{}%
\providecommand \url  [0]{\begingroup\@sanitize@url \@url }%
\providecommand \@url [1]{\endgroup\@href {#1}{\urlprefix }}%
\providecommand \urlprefix  [0]{URL }%
\providecommand \Eprint [0]{\href }%
\providecommand \doibase [0]{http://dx.doi.org/}%
\providecommand \selectlanguage [0]{\@gobble}%
\providecommand \bibinfo  [0]{\@secondoftwo}%
\providecommand \bibfield  [0]{\@secondoftwo}%
\providecommand \translation [1]{[#1]}%
\providecommand \BibitemOpen [0]{}%
\providecommand \bibitemStop [0]{}%
\providecommand \bibitemNoStop [0]{.\EOS\space}%
\providecommand \EOS [0]{\spacefactor3000\relax}%
\providecommand \BibitemShut  [1]{\csname bibitem#1\endcsname}%
\let\auto@bib@innerbib\@empty
%</preamble>
\bibitem [{\citenamefont {Welch}(2012)}]{Welch2012}%
  \BibitemOpen
  \bibfield  {author} {\bibinfo {author} {\bibfnamefont {C.}~\bibnamefont
  {Welch}},\ }\enquote {\bibinfo {title} {Frustro typeface applies the
  {P}enrose impossible triangle concept to words},}\ \ (\bibinfo  {publisher}
  {The Verge},\ \bibinfo {year} {March 23 2012})\BibitemShut {NoStop}%
\bibitem [{\citenamefont {Kumar}(2010)}]{Kumar2010}%
  \BibitemOpen
  \bibfield  {author} {\bibinfo {author} {\bibfnamefont {M.}~\bibnamefont
  {Kumar}},\ }\enquote {\bibinfo {title} {Cycles of {T}ime: {A}n
  {E}xtraordinary {N}ew {V}iew of the {U}niverse by {R}oger {P}enrose –
  review},}\ \ (\bibinfo  {publisher} {The Guardian},\ \bibinfo {year} {October
  15 2010})\BibitemShut {NoStop}%
\bibitem [{Aan(2013)}]{AandD2020}%
  \BibitemOpen
  \enquote {\bibinfo {title} {Ascending and descending by {M.C. E}scher –
  {F}acts about the {P}ainting},}\ \ (\bibinfo  {publisher} {Totally History},\
  \bibinfo {year} {May 21 2013})\BibitemShut {NoStop}%
\bibitem [{\citenamefont {Penrose}(1965)}]{PhysRevLett.14.57}%
  \BibitemOpen
  \bibfield  {author} {\bibinfo {author} {\bibfnamefont {R.}~\bibnamefont
  {Penrose}},\ }\href {\doibase 10.1103/PhysRevLett.14.57} {\bibfield
  {journal} {\bibinfo  {journal} {Phys. Rev. Lett.}\ }\textbf {\bibinfo
  {volume} {14}},\ \bibinfo {pages} {57} (\bibinfo {year} {1965})}\BibitemShut
  {NoStop}%
\bibitem [{\citenamefont {{Penrose}}(1969)}]{Penrose:1969}%
  \BibitemOpen
  \bibfield  {author} {\bibinfo {author} {\bibfnamefont {R.}~\bibnamefont
  {{Penrose}}},\ }\href@noop {} {\bibfield  {journal} {\bibinfo  {journal}
  {Nuovo Cimento Rivista Serie}\ }\textbf {\bibinfo {volume} {1}},\ \bibinfo
  {pages} {252} (\bibinfo {year} {1969})}\BibitemShut {NoStop}%
\bibitem [{\citenamefont {Penrose}(1999)}]{Penrose:1999vj}%
  \BibitemOpen
  \bibfield  {author} {\bibinfo {author} {\bibfnamefont {R.}~\bibnamefont
  {Penrose}},\ }\href {\doibase 10.1007/BF02702355} {\bibfield  {journal}
  {\bibinfo  {journal} {J. Astrophys. Astron.}\ }\textbf {\bibinfo {volume}
  {20}},\ \bibinfo {pages} {233} (\bibinfo {year} {1999})}\BibitemShut
  {NoStop}%
\bibitem [{\citenamefont {Wheeler}\ and\ \citenamefont
  {Ford}(2000)}]{WheelerBH:2000}%
  \BibitemOpen
  \bibfield  {author} {\bibinfo {author} {\bibfnamefont {J.}~\bibnamefont
  {Wheeler}}\ and\ \bibinfo {author} {\bibfnamefont {K.}~\bibnamefont {Ford}},\
  }\href@noop {} {\emph {\bibinfo {title} {Geons, Black Holes and Quantum Foam:
  A Life in Physics}}}\ (\bibinfo  {publisher} {New York: W.W. Norton \& Co.},\
  \bibinfo {address} {New York},\ \bibinfo {year} {2000})\BibitemShut {NoStop}%
\bibitem [{\citenamefont {Gannon}(1975)}]{Gannon:1975}%
  \BibitemOpen
  \bibfield  {author} {\bibinfo {author} {\bibfnamefont {D.}~\bibnamefont
  {Gannon}},\ }\href {\doibase 10.1063/1.522498} {\bibfield  {journal}
  {\bibinfo  {journal} {Journal of Mathematical Physics}\ }\textbf {\bibinfo
  {volume} {16}},\ \bibinfo {pages} {2364} (\bibinfo {year} {1975})},\ \Eprint
  {http://arxiv.org/abs/https://doi.org/10.1063/1.522498}
  {https://doi.org/10.1063/1.522498} \BibitemShut {NoStop}%
\bibitem [{\citenamefont {Friedmann}(1988)}]{Friedmann:1991}%
  \BibitemOpen
  \bibfield  {author} {\bibinfo {author} {\bibfnamefont {J.}~\bibnamefont
  {Friedmann}},\ }\enquote {\bibinfo {title} {Conceptual problems of quantum
  gravity},}\ \ (\bibinfo  {publisher} {Proceedings, Osgood Hill Conference},\
  \bibinfo {address} {Cambridge},\ \bibinfo {year} {1988})\BibitemShut
  {NoStop}%
\bibitem [{\citenamefont {Friedman}\ \emph {et~al.}(1993)\citenamefont
  {Friedman}, \citenamefont {Schleich},\ and\ \citenamefont
  {Witt}}]{Friedman:1993ty}%
  \BibitemOpen
  \bibfield  {author} {\bibinfo {author} {\bibfnamefont {J.~L.}\ \bibnamefont
  {Friedman}}, \bibinfo {author} {\bibfnamefont {K.}~\bibnamefont {Schleich}},
  \ and\ \bibinfo {author} {\bibfnamefont {D.~M.}\ \bibnamefont {Witt}},\
  }\href {\doibase 10.1103/PhysRevLett.71.1486} {\bibfield  {journal} {\bibinfo
   {journal} {Phys. Rev. Lett.}\ }\textbf {\bibinfo {volume} {71}},\ \bibinfo
  {pages} {1486} (\bibinfo {year} {1993})},\ \bibinfo {note} {[Erratum:
  Phys.Rev.Lett. 75, 1872 (1995)]},\ \Eprint
  {http://arxiv.org/abs/gr-qc/9305017} {arXiv:gr-qc/9305017} \BibitemShut
  {NoStop}%
\bibitem [{\citenamefont {Louko}\ and\ \citenamefont
  {Marolf}(1998)}]{Louko:1998dj}%
  \BibitemOpen
  \bibfield  {author} {\bibinfo {author} {\bibfnamefont {J.}~\bibnamefont
  {Louko}}\ and\ \bibinfo {author} {\bibfnamefont {D.}~\bibnamefont {Marolf}},\
  }\href {\doibase 10.1103/PhysRevD.58.024007} {\bibfield  {journal} {\bibinfo
  {journal} {Phys. Rev. D}\ }\textbf {\bibinfo {volume} {58}},\ \bibinfo
  {pages} {024007} (\bibinfo {year} {1998})},\ \Eprint
  {http://arxiv.org/abs/gr-qc/9802068} {arXiv:gr-qc/9802068} \BibitemShut
  {NoStop}%
\bibitem [{\citenamefont {Louko}\ \emph {et~al.}(2005)\citenamefont {Louko},
  \citenamefont {Mann},\ and\ \citenamefont {Marolf}}]{Louko:2004ej}%
  \BibitemOpen
  \bibfield  {author} {\bibinfo {author} {\bibfnamefont {J.}~\bibnamefont
  {Louko}}, \bibinfo {author} {\bibfnamefont {R.~B.}\ \bibnamefont {Mann}}, \
  and\ \bibinfo {author} {\bibfnamefont {D.}~\bibnamefont {Marolf}},\ }\href
  {\doibase 10.1088/0264-9381/22/7/016} {\bibfield  {journal} {\bibinfo
  {journal} {Class. Quant. Grav.}\ }\textbf {\bibinfo {volume} {22}},\ \bibinfo
  {pages} {1451} (\bibinfo {year} {2005})},\ \Eprint
  {http://arxiv.org/abs/gr-qc/0412012} {arXiv:gr-qc/0412012} \BibitemShut
  {NoStop}%
\bibitem [{\citenamefont {{Ba\~{n}ados}}\ \emph {et~al.}(1992)\citenamefont
  {{Ba\~{n}ados}}, \citenamefont {Teitelboim},\ and\ \citenamefont
  {Zanelli}}]{Banados:1992}%
  \BibitemOpen
  \bibfield  {author} {\bibinfo {author} {\bibfnamefont {M.}~\bibnamefont
  {{Ba\~{n}ados}}}, \bibinfo {author} {\bibfnamefont {C.}~\bibnamefont
  {Teitelboim}}, \ and\ \bibinfo {author} {\bibfnamefont {J.}~\bibnamefont
  {Zanelli}},\ }\href {\doibase 10.1103/PhysRevLett.69.1849} {\bibfield
  {journal} {\bibinfo  {journal} {Phys. Rev. Lett.}\ }\textbf {\bibinfo
  {volume} {69}},\ \bibinfo {pages} {1849} (\bibinfo {year}
  {1992})}\BibitemShut {NoStop}%
\bibitem [{\citenamefont {{Ba\~{n}ados}}\ \emph {et~al.}(1993)\citenamefont
  {{Ba\~{n}ados}}, \citenamefont {Henneaux}, \citenamefont {Teitelboim},\ and\
  \citenamefont {Zanelli}}]{Banados:1993}%
  \BibitemOpen
  \bibfield  {author} {\bibinfo {author} {\bibfnamefont {M.}~\bibnamefont
  {{Ba\~{n}ados}}}, \bibinfo {author} {\bibfnamefont {M.}~\bibnamefont
  {Henneaux}}, \bibinfo {author} {\bibfnamefont {C.}~\bibnamefont
  {Teitelboim}}, \ and\ \bibinfo {author} {\bibfnamefont {J.}~\bibnamefont
  {Zanelli}},\ }\href {\doibase 10.1103/PhysRevD.48.1506} {\bibfield  {journal}
  {\bibinfo  {journal} {Phys. Rev. D}\ }\textbf {\bibinfo {volume} {48}},\
  \bibinfo {pages} {1506} (\bibinfo {year} {1993})}\BibitemShut {NoStop}%
\bibitem [{\citenamefont {Carlip}(2003)}]{Carlip:2003}%
  \BibitemOpen
  \bibfield  {author} {\bibinfo {author} {\bibfnamefont {S.}~\bibnamefont
  {Carlip}},\ }\href@noop {} {\emph {\bibinfo {title} {Quantum Gravity in 2+1
  Dimensions}}}\ (\bibinfo  {publisher} {Cambridge University Press},\ \bibinfo
  {address} {Cambridge},\ \bibinfo {year} {2003})\BibitemShut {NoStop}%
\bibitem [{\citenamefont {Louko}\ and\ \citenamefont
  {Marolf}(1999)}]{Louko:1998hc}%
  \BibitemOpen
  \bibfield  {author} {\bibinfo {author} {\bibfnamefont {J.}~\bibnamefont
  {Louko}}\ and\ \bibinfo {author} {\bibfnamefont {D.}~\bibnamefont {Marolf}},\
  }\href {\doibase 10.1103/PhysRevD.59.066002} {\bibfield  {journal} {\bibinfo
  {journal} {Phys. Rev. D}\ }\textbf {\bibinfo {volume} {59}},\ \bibinfo
  {pages} {066002} (\bibinfo {year} {1999})},\ \Eprint
  {http://arxiv.org/abs/hep-th/9808081} {arXiv:hep-th/9808081} \BibitemShut
  {NoStop}%
\bibitem [{\citenamefont {Louko}\ \emph {et~al.}(2000)\citenamefont {Louko},
  \citenamefont {Marolf},\ and\ \citenamefont {Ross}}]{Louko:2000tp}%
  \BibitemOpen
  \bibfield  {author} {\bibinfo {author} {\bibfnamefont {J.}~\bibnamefont
  {Louko}}, \bibinfo {author} {\bibfnamefont {D.}~\bibnamefont {Marolf}}, \
  and\ \bibinfo {author} {\bibfnamefont {S.~F.}\ \bibnamefont {Ross}},\ }\href
  {\doibase 10.1103/PhysRevD.62.044041} {\bibfield  {journal} {\bibinfo
  {journal} {Phys. Rev. D}\ }\textbf {\bibinfo {volume} {62}},\ \bibinfo
  {pages} {044041} (\bibinfo {year} {2000})},\ \Eprint
  {http://arxiv.org/abs/hep-th/0002111} {arXiv:hep-th/0002111} \BibitemShut
  {NoStop}%
\bibitem [{\citenamefont {Maldacena}(2003)}]{Maldacena:2001kr}%
  \BibitemOpen
  \bibfield  {author} {\bibinfo {author} {\bibfnamefont {J.~M.}\ \bibnamefont
  {Maldacena}},\ }\href {\doibase 10.1088/1126-6708/2003/04/021} {\bibfield
  {journal} {\bibinfo  {journal} {JHEP}\ }\textbf {\bibinfo {volume} {04}},\
  \bibinfo {pages} {021} (\bibinfo {year} {2003})},\ \Eprint
  {http://arxiv.org/abs/hep-th/0106112} {arXiv:hep-th/0106112} \BibitemShut
  {NoStop}%
\bibitem [{\citenamefont {Sinamuli}\ and\ \citenamefont
  {Mann}(2017)}]{Sinamuli:2016rms}%
  \BibitemOpen
  \bibfield  {author} {\bibinfo {author} {\bibfnamefont {M.}~\bibnamefont
  {Sinamuli}}\ and\ \bibinfo {author} {\bibfnamefont {R.~B.}\ \bibnamefont
  {Mann}},\ }\href {\doibase 10.1103/PhysRevD.96.026014} {\bibfield  {journal}
  {\bibinfo  {journal} {Phys. Rev. D}\ }\textbf {\bibinfo {volume} {96}},\
  \bibinfo {pages} {026014} (\bibinfo {year} {2017})},\ \Eprint
  {http://arxiv.org/abs/1612.06880} {arXiv:1612.06880 [hep-th]} \BibitemShut
  {NoStop}%
\bibitem [{\citenamefont {Sinamuli}\ and\ \citenamefont
  {Mann}(2018)}]{Sinamuli:2018jhm}%
  \BibitemOpen
  \bibfield  {author} {\bibinfo {author} {\bibfnamefont {M.}~\bibnamefont
  {Sinamuli}}\ and\ \bibinfo {author} {\bibfnamefont {R.~B.}\ \bibnamefont
  {Mann}},\ }\href {\doibase 10.1103/PhysRevD.98.026005} {\bibfield  {journal}
  {\bibinfo  {journal} {Phys. Rev. D}\ }\textbf {\bibinfo {volume} {98}},\
  \bibinfo {pages} {026005} (\bibinfo {year} {2018})},\ \Eprint
  {http://arxiv.org/abs/1804.07333} {arXiv:1804.07333 [hep-th]} \BibitemShut
  {NoStop}%
\bibitem [{\citenamefont {Miyaji}\ \emph {et~al.}(2015)\citenamefont {Miyaji},
  \citenamefont {Numasawa}, \citenamefont {Shiba}, \citenamefont {Takayanagi},\
  and\ \citenamefont {Watanabe}}]{Miyaji:2015woj}%
  \BibitemOpen
  \bibfield  {author} {\bibinfo {author} {\bibfnamefont {M.}~\bibnamefont
  {Miyaji}}, \bibinfo {author} {\bibfnamefont {T.}~\bibnamefont {Numasawa}},
  \bibinfo {author} {\bibfnamefont {N.}~\bibnamefont {Shiba}}, \bibinfo
  {author} {\bibfnamefont {T.}~\bibnamefont {Takayanagi}}, \ and\ \bibinfo
  {author} {\bibfnamefont {K.}~\bibnamefont {Watanabe}},\ }\href {\doibase
  10.1103/PhysRevLett.115.261602} {\bibfield  {journal} {\bibinfo  {journal}
  {Phys. Rev. Lett.}\ }\textbf {\bibinfo {volume} {115}},\ \bibinfo {pages}
  {261602} (\bibinfo {year} {2015})},\ \Eprint
  {http://arxiv.org/abs/1507.07555} {arXiv:1507.07555 [hep-th]} \BibitemShut
  {NoStop}%
\bibitem [{\citenamefont {Smith}\ and\ \citenamefont
  {Mann}(2014)}]{Smith:2014}%
  \BibitemOpen
  \bibfield  {author} {\bibinfo {author} {\bibfnamefont {A.~R.~H.}\
  \bibnamefont {Smith}}\ and\ \bibinfo {author} {\bibfnamefont {R.~B.}\
  \bibnamefont {Mann}},\ }\href {\doibase 10.1088/0264-9381/31/8/082001}
  {\bibfield  {journal} {\bibinfo  {journal} {Class. Quant. Grav.}\ }\textbf
  {\bibinfo {volume} {31}},\ \bibinfo {pages} {082001} (\bibinfo {year}
  {2014})}\BibitemShut {NoStop}%
\bibitem [{\citenamefont {Ng}\ \emph {et~al.}(2017)\citenamefont {Ng},
  \citenamefont {Mann},\ and\ \citenamefont
  {Mart\'{\i}n-Mart\'{\i}nez}}]{Ng2017}%
  \BibitemOpen
  \bibfield  {author} {\bibinfo {author} {\bibfnamefont {K.~K.}\ \bibnamefont
  {Ng}}, \bibinfo {author} {\bibfnamefont {R.~B.}\ \bibnamefont {Mann}}, \ and\
  \bibinfo {author} {\bibfnamefont {E.}~\bibnamefont
  {Mart\'{\i}n-Mart\'{\i}nez}},\ }\href {\doibase 10.1103/PhysRevD.96.085004}
  {\bibfield  {journal} {\bibinfo  {journal} {Phys. Rev. D}\ }\textbf {\bibinfo
  {volume} {96}},\ \bibinfo {pages} {085004} (\bibinfo {year}
  {2017})}\BibitemShut {NoStop}%
\bibitem [{\citenamefont {Unruh}(1976)}]{Unruh1976}%
  \BibitemOpen
  \bibfield  {author} {\bibinfo {author} {\bibfnamefont {W.~G.}\ \bibnamefont
  {Unruh}},\ }\href {\doibase 10.1103/PhysRevD.14.870} {\bibfield  {journal}
  {\bibinfo  {journal} {Phys. Rev. D}\ }\textbf {\bibinfo {volume} {14}},\
  \bibinfo {pages} {870} (\bibinfo {year} {1976})}\BibitemShut {NoStop}%
\bibitem [{\citenamefont {DeWitt}(1979)}]{deWitt}%
  \BibitemOpen
  \bibfield  {author} {\bibinfo {author} {\bibfnamefont {B.~S.}\ \bibnamefont
  {DeWitt}},\ }in\ \href@noop {} {\emph {\bibinfo {booktitle} {General
  Relativity: An Einstein Centenary Surve}}}\ (\bibinfo  {publisher} {Cambridge
  University Press},\ \bibinfo {address} {Cambridge},\ \bibinfo {year} {1979})\
  pp.\ \bibinfo {pages} {680--745}\BibitemShut {NoStop}%
\bibitem [{\citenamefont {Henderson}\ \emph {et~al.}(2018)\citenamefont
  {Henderson}, \citenamefont {Hennigar}, \citenamefont {Mann}, \citenamefont
  {Smith},\ and\ \citenamefont {Zhang}}]{henderson:2018}%
  \BibitemOpen
  \bibfield  {author} {\bibinfo {author} {\bibfnamefont {L.~J.}\ \bibnamefont
  {Henderson}}, \bibinfo {author} {\bibfnamefont {R.~A.}\ \bibnamefont
  {Hennigar}}, \bibinfo {author} {\bibfnamefont {R.~B.}\ \bibnamefont {Mann}},
  \bibinfo {author} {\bibfnamefont {A.~R.~H.}\ \bibnamefont {Smith}}, \ and\
  \bibinfo {author} {\bibfnamefont {J.}~\bibnamefont {Zhang}},\ }\href
  {\doibase 10.1088/1361-6382/aae27e} {\bibfield  {journal} {\bibinfo
  {journal} {Classical and Quantum Gravity}\ }\textbf {\bibinfo {volume}
  {35}},\ \bibinfo {pages} {21LT02} (\bibinfo {year} {2018})}\BibitemShut
  {NoStop}%
\bibitem [{\citenamefont {Henderson}\ \emph {et~al.}(2020)\citenamefont
  {Henderson}, \citenamefont {Hennigar}, \citenamefont {Mann}, \citenamefont
  {Smith},\ and\ \citenamefont {Zhang}}]{AH:plb2020}%
  \BibitemOpen
  \bibfield  {author} {\bibinfo {author} {\bibfnamefont {L.}~\bibnamefont
  {Henderson}}, \bibinfo {author} {\bibfnamefont {R.}~\bibnamefont {Hennigar}},
  \bibinfo {author} {\bibfnamefont {R.}~\bibnamefont {Mann}}, \bibinfo {author}
  {\bibfnamefont {A.}~\bibnamefont {Smith}}, \ and\ \bibinfo {author}
  {\bibfnamefont {J.-L.}\ \bibnamefont {Zhang}},\ }\href {\doibase
  10.1016/j.physletb.2020.135732} {\bibfield  {journal} {\bibinfo  {journal}
  {Physics Letters B}\ }\textbf {\bibinfo {volume} {809}},\ \bibinfo {pages}
  {135732} (\bibinfo {year} {2020})}\BibitemShut {NoStop}%
\bibitem [{\citenamefont {Robbins}\ \emph {et~al.}(2020)\citenamefont
  {Robbins}, \citenamefont {Henderson},\ and\ \citenamefont
  {Mann}}]{Robbins:2020jca}%
  \BibitemOpen
  \bibfield  {author} {\bibinfo {author} {\bibfnamefont {M.~P.~G.}\
  \bibnamefont {Robbins}}, \bibinfo {author} {\bibfnamefont {L.~J.}\
  \bibnamefont {Henderson}}, \ and\ \bibinfo {author} {\bibfnamefont {R.~B.}\
  \bibnamefont {Mann}},\ }\href@noop {} {\  (\bibinfo {year} {2020})},\ \Eprint
  {http://arxiv.org/abs/2010.14517} {arXiv:2010.14517 [hep-th]} \BibitemShut
  {NoStop}%
\bibitem [{\citenamefont {Peres}\ and\ \citenamefont
  {Terno}(2004)}]{Peres2004}%
  \BibitemOpen
  \bibfield  {author} {\bibinfo {author} {\bibfnamefont {A.}~\bibnamefont
  {Peres}}\ and\ \bibinfo {author} {\bibfnamefont {D.~R.}\ \bibnamefont
  {Terno}},\ }\href {\doibase 10.1103/RevModPhys.76.93} {\bibfield  {journal}
  {\bibinfo  {journal} {Rev. Mod. Phys.}\ }\textbf {\bibinfo {volume} {76}},\
  \bibinfo {pages} {93} (\bibinfo {year} {2004})}\BibitemShut {NoStop}%
\bibitem [{\citenamefont {Lamata}\ \emph {et~al.}(2006)\citenamefont {Lamata},
  \citenamefont {Martin-Delgado},\ and\ \citenamefont {Solano}}]{Lamata1997}%
  \BibitemOpen
  \bibfield  {author} {\bibinfo {author} {\bibfnamefont {L.}~\bibnamefont
  {Lamata}}, \bibinfo {author} {\bibfnamefont {M.~A.}\ \bibnamefont
  {Martin-Delgado}}, \ and\ \bibinfo {author} {\bibfnamefont {E.}~\bibnamefont
  {Solano}},\ }\href {\doibase 10.1103/PhysRevLett.97.250502} {\bibfield
  {journal} {\bibinfo  {journal} {Phys. Rev. Lett.}\ }\textbf {\bibinfo
  {volume} {97}},\ \bibinfo {pages} {250502} (\bibinfo {year}
  {2006})}\BibitemShut {NoStop}%
\bibitem [{\citenamefont {Ralph}\ \emph {et~al.}(2009)\citenamefont {Ralph},
  \citenamefont {Milburn},\ and\ \citenamefont {Downes}}]{Ralph2009}%
  \BibitemOpen
  \bibfield  {author} {\bibinfo {author} {\bibfnamefont {T.~C.}\ \bibnamefont
  {Ralph}}, \bibinfo {author} {\bibfnamefont {G.~J.}\ \bibnamefont {Milburn}},
  \ and\ \bibinfo {author} {\bibfnamefont {T.}~\bibnamefont {Downes}},\ }\href
  {\doibase 10.1103/PhysRevA.79.022121} {\bibfield  {journal} {\bibinfo
  {journal} {Phys. Rev. A}\ }\textbf {\bibinfo {volume} {79}},\ \bibinfo
  {pages} {022121} (\bibinfo {year} {2009})}\BibitemShut {NoStop}%
\bibitem [{\citenamefont {Ryu}\ and\ \citenamefont
  {Takayanagi}(2006)}]{Ryu2006}%
  \BibitemOpen
  \bibfield  {author} {\bibinfo {author} {\bibfnamefont {S.}~\bibnamefont
  {Ryu}}\ and\ \bibinfo {author} {\bibfnamefont {T.}~\bibnamefont
  {Takayanagi}},\ }\href {\doibase 10.1103/PhysRevLett.96.181602} {\bibfield
  {journal} {\bibinfo  {journal} {Phys. Rev. Lett.}\ }\textbf {\bibinfo
  {volume} {96}},\ \bibinfo {pages} {181602} (\bibinfo {year} {2006})},\
  \Eprint {http://arxiv.org/abs/hep-th/0603001} {arXiv:hep-th/0603001}
  \BibitemShut {NoStop}%
\bibitem [{\citenamefont {Hotta}(2009)}]{doi:10.1143/JPSJ.78.034001}%
  \BibitemOpen
  \bibfield  {author} {\bibinfo {author} {\bibfnamefont {M.}~\bibnamefont
  {Hotta}},\ }\href {\doibase 10.1143/JPSJ.78.034001} {\bibfield  {journal}
  {\bibinfo  {journal} {Journal of the Physical Society of Japan}\ }\textbf
  {\bibinfo {volume} {78}},\ \bibinfo {pages} {034001} (\bibinfo {year}
  {2009})},\ \Eprint
  {http://arxiv.org/abs/https://doi.org/10.1143/JPSJ.78.034001}
  {https://doi.org/10.1143/JPSJ.78.034001} \BibitemShut {NoStop}%
\bibitem [{\citenamefont {Hotta}(2011)}]{Hotta:2011xj}%
  \BibitemOpen
  \bibfield  {author} {\bibinfo {author} {\bibfnamefont {M.}~\bibnamefont
  {Hotta}},\ }\href@noop {} {\bibfield  {journal} {\bibinfo  {journal} {arXiv
  preprint arXiv:1101.3954}\ } (\bibinfo {year} {2011})}\BibitemShut {NoStop}%
\bibitem [{\citenamefont {{Solodukhin}}(2011)}]{Solodukhin2011}%
  \BibitemOpen
  \bibfield  {author} {\bibinfo {author} {\bibfnamefont {S.~N.}\ \bibnamefont
  {{Solodukhin}}},\ }\href {\doibase 10.12942/lrr-2011-8} {\bibfield  {journal}
  {\bibinfo  {journal} {Living Reviews in Relativity}\ }\textbf {\bibinfo
  {volume} {14}},\ \bibinfo {eid} {8} (\bibinfo {year} {2011})},\ \Eprint
  {http://arxiv.org/abs/1104.3712} {arXiv:1104.3712 [hep-th]} \BibitemShut
  {NoStop}%
\bibitem [{\citenamefont {Brustein}\ \emph {et~al.}(2006)\citenamefont
  {Brustein}, \citenamefont {Einhorn},\ and\ \citenamefont
  {Yarom}}]{Brustein2005}%
  \BibitemOpen
  \bibfield  {author} {\bibinfo {author} {\bibfnamefont {R.}~\bibnamefont
  {Brustein}}, \bibinfo {author} {\bibfnamefont {M.~B.}\ \bibnamefont
  {Einhorn}}, \ and\ \bibinfo {author} {\bibfnamefont {A.}~\bibnamefont
  {Yarom}},\ }\href {\doibase 10.1088/1126-6708/2006/01/098} {\bibfield
  {journal} {\bibinfo  {journal} {JHEP}\ }\textbf {\bibinfo {volume} {01}},\
  \bibinfo {pages} {098} (\bibinfo {year} {2006})},\ \Eprint
  {http://arxiv.org/abs/hep-th/0508217} {arXiv:hep-th/0508217} \BibitemShut
  {NoStop}%
\bibitem [{\citenamefont {Preskill}(1992)}]{Preskill:1992tc}%
  \BibitemOpen
  \bibfield  {author} {\bibinfo {author} {\bibfnamefont {J.}~\bibnamefont
  {Preskill}},\ }in\ \href@noop {} {\emph {\bibinfo {booktitle} {{International
  Symposium on Black holes, Membranes, Wormholes and Superstrings Woodlands,
  Texas, January 16-18, 1992}}}}\ (\bibinfo {year} {1992})\ pp.\ \bibinfo
  {pages} {22--39},\ \Eprint {http://arxiv.org/abs/hep-th/9209058}
  {arXiv:hep-th/9209058 [hep-th]} \BibitemShut {NoStop}%
%%CITATION = HEP-TH/9209058;%%
\bibitem [{\citenamefont {Mathur}(2009)}]{Mathur:2009hf}%
  \BibitemOpen
  \bibfield  {author} {\bibinfo {author} {\bibfnamefont {S.~D.}\ \bibnamefont
  {Mathur}},\ }\href {\doibase 10.1088/0264-9381/26/22/224001} {\bibfield
  {journal} {\bibinfo  {journal} {Class.Quant.Grav.}\ }\textbf {\bibinfo
  {volume} {26}},\ \bibinfo {pages} {224001} (\bibinfo {year} {2009})},\
  \Eprint {http://arxiv.org/abs/0909.1038} {arXiv:0909.1038 [hep-th]}
  \BibitemShut {NoStop}%
%%CITATION = ARXIV:0909.1038;%%
\bibitem [{\citenamefont {Almheiri}\ \emph {et~al.}(2013)\citenamefont
  {Almheiri}, \citenamefont {Marolf}, \citenamefont {Polchinski},\ and\
  \citenamefont {Sully}}]{Almheiri:2012rt}%
  \BibitemOpen
  \bibfield  {author} {\bibinfo {author} {\bibfnamefont {A.}~\bibnamefont
  {Almheiri}}, \bibinfo {author} {\bibfnamefont {D.}~\bibnamefont {Marolf}},
  \bibinfo {author} {\bibfnamefont {J.}~\bibnamefont {Polchinski}}, \ and\
  \bibinfo {author} {\bibfnamefont {J.}~\bibnamefont {Sully}},\ }\href
  {\doibase 10.1007/JHEP02(2013)062} {\bibfield  {journal} {\bibinfo  {journal}
  {JHEP}\ }\textbf {\bibinfo {volume} {02}},\ \bibinfo {pages} {062} (\bibinfo
  {year} {2013})},\ \Eprint {http://arxiv.org/abs/1207.3123} {arXiv:1207.3123
  [hep-th]} \BibitemShut {NoStop}%
%%CITATION = ARXIV:1207.3123;%%
\bibitem [{\citenamefont {Braunstein}\ \emph {et~al.}(2013)\citenamefont
  {Braunstein}, \citenamefont {Pirandola},\ and\ \citenamefont
  {Zyczkowski}}]{Braunstein:2009my}%
  \BibitemOpen
  \bibfield  {author} {\bibinfo {author} {\bibfnamefont {S.~L.}\ \bibnamefont
  {Braunstein}}, \bibinfo {author} {\bibfnamefont {S.}~\bibnamefont
  {Pirandola}}, \ and\ \bibinfo {author} {\bibfnamefont {K.}~\bibnamefont
  {Zyczkowski}},\ }\href {\doibase 10.1103/PhysRevLett.110.101301} {\bibfield
  {journal} {\bibinfo  {journal} {Phys. Rev. Lett.}\ }\textbf {\bibinfo
  {volume} {110}},\ \bibinfo {pages} {101301} (\bibinfo {year} {2013})},\
  \Eprint {http://arxiv.org/abs/0907.1190} {arXiv:0907.1190 [quant-ph]}
  \BibitemShut {NoStop}%
%%CITATION = ARXIV:0907.1190;%%
\bibitem [{\citenamefont {Mann}(2015)}]{Mann:2015luq}%
  \BibitemOpen
  \bibfield  {author} {\bibinfo {author} {\bibfnamefont {R.~B.}\ \bibnamefont
  {Mann}},\ }\href {\doibase 10.1007/978-3-319-14496-2} {\emph {\bibinfo
  {title} {{Black Holes: Thermodynamics, Information, and Firewalls}}}},\
  SpringerBriefs in Physics\ (\bibinfo  {publisher} {Springer},\ \bibinfo
  {year} {2015})\BibitemShut {NoStop}%
%%CITATION = INSPIRE-1453974;%%
\bibitem [{\citenamefont {Summers}\ and\ \citenamefont
  {Werner}(1985)}]{summers1985bell}%
  \BibitemOpen
  \bibfield  {author} {\bibinfo {author} {\bibfnamefont {S.~J.}\ \bibnamefont
  {Summers}}\ and\ \bibinfo {author} {\bibfnamefont {R.}~\bibnamefont
  {Werner}},\ }\href {\doibase https://doi.org/10.1016/0375-9601(85)90093-3}
  {\bibfield  {journal} {\bibinfo  {journal} {Physics Letters A}\ }\textbf
  {\bibinfo {volume} {110}},\ \bibinfo {pages} {257 } (\bibinfo {year}
  {1985})}\BibitemShut {NoStop}%
\bibitem [{\citenamefont {Summers}\ and\ \citenamefont
  {Werner}()}]{summers_bells_1987}%
  \BibitemOpen
  \bibfield  {author} {\bibinfo {author} {\bibfnamefont {S.~J.}\ \bibnamefont
  {Summers}}\ and\ \bibinfo {author} {\bibfnamefont {R.}~\bibnamefont
  {Werner}},\ }\href {\doibase 10.1063/1.527734} {\ \textbf {\bibinfo {volume}
  {28}},\ \bibinfo {pages} {2448}}\BibitemShut {NoStop}%
\bibitem [{\citenamefont {Valentini}(1991)}]{Valentini1991}%
  \BibitemOpen
  \bibfield  {author} {\bibinfo {author} {\bibfnamefont {A.}~\bibnamefont
  {Valentini}},\ }\href {\doibase https://doi.org/10.1016/0375-9601(91)90952-5}
  {\bibfield  {journal} {\bibinfo  {journal} {Physics Letters A}\ }\textbf
  {\bibinfo {volume} {153}},\ \bibinfo {pages} {321 } (\bibinfo {year}
  {1991})}\BibitemShut {NoStop}%
\bibitem [{\citenamefont {Reznik}(2003)}]{Reznik2003}%
  \BibitemOpen
  \bibfield  {author} {\bibinfo {author} {\bibfnamefont {B.}~\bibnamefont
  {Reznik}},\ }\href@noop {} {\bibfield  {journal} {\bibinfo  {journal}
  {Foundations of Physics}\ }\textbf {\bibinfo {volume} {33}},\ \bibinfo
  {pages} {167} (\bibinfo {year} {2003})}\BibitemShut {NoStop}%
\bibitem [{\citenamefont {Reznik}\ \emph {et~al.}(2005)\citenamefont {Reznik},
  \citenamefont {Retzker},\ and\ \citenamefont {Silman}}]{Reznik2005}%
  \BibitemOpen
  \bibfield  {author} {\bibinfo {author} {\bibfnamefont {B.}~\bibnamefont
  {Reznik}}, \bibinfo {author} {\bibfnamefont {A.}~\bibnamefont {Retzker}}, \
  and\ \bibinfo {author} {\bibfnamefont {J.}~\bibnamefont {Silman}},\ }\href
  {\doibase 10.1103/PhysRevA.71.042104} {\bibfield  {journal} {\bibinfo
  {journal} {Phys. Rev. A}\ }\textbf {\bibinfo {volume} {71}},\ \bibinfo
  {pages} {042104} (\bibinfo {year} {2005})}\BibitemShut {NoStop}%
\bibitem [{\citenamefont {Salton}\ \emph {et~al.}(2015)\citenamefont {Salton},
  \citenamefont {Mann},\ and\ \citenamefont {Menicucci}}]{Salton2015}%
  \BibitemOpen
  \bibfield  {author} {\bibinfo {author} {\bibfnamefont {G.}~\bibnamefont
  {Salton}}, \bibinfo {author} {\bibfnamefont {R.~B.}\ \bibnamefont {Mann}}, \
  and\ \bibinfo {author} {\bibfnamefont {N.~C.}\ \bibnamefont {Menicucci}},\
  }\href {\doibase 10.1088/1367-2630/17/3/035001} {\bibfield  {journal}
  {\bibinfo  {journal} {New Journal of Physics}\ }\textbf {\bibinfo {volume}
  {17}},\ \bibinfo {pages} {035001} (\bibinfo {year} {2015})}\BibitemShut
  {NoStop}%
\bibitem [{\citenamefont {Mart\'{\i}n-Mart\'{\i}nez}\ \emph
  {et~al.}(2016)\citenamefont {Mart\'{\i}n-Mart\'{\i}nez}, \citenamefont
  {Smith},\ and\ \citenamefont {Terno}}]{Smith2016}%
  \BibitemOpen
  \bibfield  {author} {\bibinfo {author} {\bibfnamefont {E.}~\bibnamefont
  {Mart\'{\i}n-Mart\'{\i}nez}}, \bibinfo {author} {\bibfnamefont {A.~R.~H.}\
  \bibnamefont {Smith}}, \ and\ \bibinfo {author} {\bibfnamefont {D.~R.}\
  \bibnamefont {Terno}},\ }\href {\doibase 10.1103/PhysRevD.93.044001}
  {\bibfield  {journal} {\bibinfo  {journal} {Phys. Rev. D}\ }\textbf {\bibinfo
  {volume} {93}},\ \bibinfo {pages} {044001} (\bibinfo {year}
  {2016})}\BibitemShut {NoStop}%
\bibitem [{\citenamefont {Steeg}\ and\ \citenamefont
  {Menicucci}(2009)}]{PhysRevD.79.044027}%
  \BibitemOpen
  \bibfield  {author} {\bibinfo {author} {\bibfnamefont {G.~V.}\ \bibnamefont
  {Steeg}}\ and\ \bibinfo {author} {\bibfnamefont {N.~C.}\ \bibnamefont
  {Menicucci}},\ }\href {\doibase 10.1103/PhysRevD.79.044027} {\bibfield
  {journal} {\bibinfo  {journal} {Phys. Rev. D}\ }\textbf {\bibinfo {volume}
  {79}},\ \bibinfo {pages} {044027} (\bibinfo {year} {2009})}\BibitemShut
  {NoStop}%
\bibitem [{\citenamefont {Huang}\ and\ \citenamefont {Tian}(2017)}]{Huang2017}%
  \BibitemOpen
  \bibfield  {author} {\bibinfo {author} {\bibfnamefont {Z.}~\bibnamefont
  {Huang}}\ and\ \bibinfo {author} {\bibfnamefont {Z.}~\bibnamefont {Tian}},\
  }\href {\doibase https://doi.org/10.1016/j.nuclphysb.2017.08.014} {\bibfield
  {journal} {\bibinfo  {journal} {Nuclear Physics B}\ }\textbf {\bibinfo
  {volume} {923}},\ \bibinfo {pages} {458 } (\bibinfo {year}
  {2017})}\BibitemShut {NoStop}%
\bibitem [{\citenamefont {Ng}\ \emph {et~al.}(2018)\citenamefont {Ng},
  \citenamefont {Mann},\ and\ \citenamefont {Martín-Martínez}}]{Ng:2018drz}%
  \BibitemOpen
  \bibfield  {author} {\bibinfo {author} {\bibfnamefont {K.~K.}\ \bibnamefont
  {Ng}}, \bibinfo {author} {\bibfnamefont {R.~B.}\ \bibnamefont {Mann}}, \ and\
  \bibinfo {author} {\bibfnamefont {E.}~\bibnamefont {Martín-Martínez}},\
  }\href {\doibase 10.1103/PhysRevD.98.125005} {\bibfield  {journal} {\bibinfo
  {journal} {Phys. Rev. D}\ }\textbf {\bibinfo {volume} {98}},\ \bibinfo
  {pages} {125005} (\bibinfo {year} {2018})},\ \Eprint
  {http://arxiv.org/abs/1809.06878} {arXiv:1809.06878 [quant-ph]} \BibitemShut
  {NoStop}%
\bibitem [{\citenamefont {Henderson}\ \emph
  {et~al.}(2019{\natexlab{a}})\citenamefont {Henderson}, \citenamefont
  {Hennigar}, \citenamefont {Mann}, \citenamefont {Smith},\ and\ \citenamefont
  {Zhang}}]{Henderson2019}%
  \BibitemOpen
  \bibfield  {author} {\bibinfo {author} {\bibfnamefont {L.~J.}\ \bibnamefont
  {Henderson}}, \bibinfo {author} {\bibfnamefont {R.~A.}\ \bibnamefont
  {Hennigar}}, \bibinfo {author} {\bibfnamefont {R.~B.}\ \bibnamefont {Mann}},
  \bibinfo {author} {\bibfnamefont {A.~R.~H.}\ \bibnamefont {Smith}}, \ and\
  \bibinfo {author} {\bibfnamefont {J.}~\bibnamefont {Zhang}},\ }\href
  {\doibase 10.1007/jhep05(2019)178} {\bibfield  {journal} {\bibinfo  {journal}
  {Journal of High Energy Physics}\ }\textbf {\bibinfo {volume} {2019}},\
  \bibinfo {pages} {178} (\bibinfo {year} {2019}{\natexlab{a}})}\BibitemShut
  {NoStop}%
\bibitem [{\citenamefont {Ba\~nados}\ \emph {et~al.}(1992)\citenamefont
  {Ba\~nados}, \citenamefont {Teitelboim},\ and\ \citenamefont
  {Zanelli}}]{Banados1992}%
  \BibitemOpen
  \bibfield  {author} {\bibinfo {author} {\bibfnamefont {M.}~\bibnamefont
  {Ba\~nados}}, \bibinfo {author} {\bibfnamefont {C.}~\bibnamefont
  {Teitelboim}}, \ and\ \bibinfo {author} {\bibfnamefont {J.}~\bibnamefont
  {Zanelli}},\ }\href {\doibase 10.1103/PhysRevLett.69.1849} {\bibfield
  {journal} {\bibinfo  {journal} {Phys. Rev. Lett.}\ }\textbf {\bibinfo
  {volume} {69}},\ \bibinfo {pages} {1849} (\bibinfo {year}
  {1992})}\BibitemShut {NoStop}%
\bibitem [{\citenamefont {Henderson}\ \emph
  {et~al.}(2019{\natexlab{b}})\citenamefont {Henderson}, \citenamefont
  {Hennigar}, \citenamefont {Mann}, \citenamefont {Smith},\ and\ \citenamefont
  {Zhang}}]{Henderson:2019uqo}%
  \BibitemOpen
  \bibfield  {author} {\bibinfo {author} {\bibfnamefont {L.~J.}\ \bibnamefont
  {Henderson}}, \bibinfo {author} {\bibfnamefont {R.~A.}\ \bibnamefont
  {Hennigar}}, \bibinfo {author} {\bibfnamefont {R.~B.}\ \bibnamefont {Mann}},
  \bibinfo {author} {\bibfnamefont {A.~R.}\ \bibnamefont {Smith}}, \ and\
  \bibinfo {author} {\bibfnamefont {J.}~\bibnamefont {Zhang}},\ }\href@noop {}
  {\  (\bibinfo {year} {2019}{\natexlab{b}})},\ \Eprint
  {http://arxiv.org/abs/1911.02977} {arXiv:1911.02977 [gr-qc]} \BibitemShut
  {NoStop}%
\bibitem [{\citenamefont {{Tjoa}}\ and\ \citenamefont
  {{Mann}}(2020)}]{Tjoa2020}%
  \BibitemOpen
  \bibfield  {author} {\bibinfo {author} {\bibfnamefont {E.}~\bibnamefont
  {{Tjoa}}}\ and\ \bibinfo {author} {\bibfnamefont {R.~B.}\ \bibnamefont
  {{Mann}}},\ }\href@noop {} {\bibfield  {journal} {\bibinfo  {journal} {arXiv
  e-prints}\ ,\ \bibinfo {eid} {arXiv:2007.02955}} (\bibinfo {year} {2020})},\
  \Eprint {http://arxiv.org/abs/2007.02955} {arXiv:2007.02955 [quant-ph]}
  \BibitemShut {NoStop}%
\bibitem [{\citenamefont {Smith}(2017)}]{Smith:2017vle}%
  \BibitemOpen
  \bibfield  {author} {\bibinfo {author} {\bibfnamefont {A.~R.}\ \bibnamefont
  {Smith}},\ }\emph {\bibinfo {title} {{Detectors, Reference Frames, and
  Time}}},\ \href {\doibase 10.1007/978-3-030-11000-0} {Ph.D. thesis},\
  \bibinfo  {school} {U. Waterloo (main)} (\bibinfo {year} {2017})\BibitemShut
  {NoStop}%
\bibitem [{\citenamefont {Wootters}(1998)}]{Wooters1998}%
  \BibitemOpen
  \bibfield  {author} {\bibinfo {author} {\bibfnamefont {W.~K.}\ \bibnamefont
  {Wootters}},\ }\href {\doibase 10.1103/PhysRevLett.80.2245} {\bibfield
  {journal} {\bibinfo  {journal} {Phys. Rev. Lett.}\ }\textbf {\bibinfo
  {volume} {80}},\ \bibinfo {pages} {2245} (\bibinfo {year}
  {1998})}\BibitemShut {NoStop}%
\bibitem [{\citenamefont {Mart\'{\i}n-Mart\'{\i}nez}\ \emph
  {et~al.}(2013)\citenamefont {Mart\'{\i}n-Mart\'{\i}nez}, \citenamefont
  {Montero},\ and\ \citenamefont {del Rey}}]{Martin-Martinez2013}%
  \BibitemOpen
  \bibfield  {author} {\bibinfo {author} {\bibfnamefont {E.}~\bibnamefont
  {Mart\'{\i}n-Mart\'{\i}nez}}, \bibinfo {author} {\bibfnamefont
  {M.}~\bibnamefont {Montero}}, \ and\ \bibinfo {author} {\bibfnamefont
  {M.}~\bibnamefont {del Rey}},\ }\href {\doibase 10.1103/PhysRevD.87.064038}
  {\bibfield  {journal} {\bibinfo  {journal} {Phys. Rev. D}\ }\textbf {\bibinfo
  {volume} {87}},\ \bibinfo {pages} {064038} (\bibinfo {year}
  {2013})}\BibitemShut {NoStop}%
\bibitem [{\citenamefont {Alhambra}\ \emph {et~al.}(2014)\citenamefont
  {Alhambra}, \citenamefont {Kempf},\ and\ \citenamefont
  {Mart\'{\i}n-Mart\'{\i}nez}}]{Alvaro}%
  \BibitemOpen
  \bibfield  {author} {\bibinfo {author} {\bibfnamefont {A.~M.}\ \bibnamefont
  {Alhambra}}, \bibinfo {author} {\bibfnamefont {A.}~\bibnamefont {Kempf}}, \
  and\ \bibinfo {author} {\bibfnamefont {E.}~\bibnamefont
  {Mart\'{\i}n-Mart\'{\i}nez}},\ }\href {\doibase 10.1103/PhysRevA.89.033835}
  {\bibfield  {journal} {\bibinfo  {journal} {Phys. Rev. A}\ }\textbf {\bibinfo
  {volume} {89}},\ \bibinfo {pages} {033835} (\bibinfo {year}
  {2014})}\BibitemShut {NoStop}%
\bibitem [{\citenamefont {Pozas-Kerstjens}\ and\ \citenamefont
  {Mart\'{i}n-Mart\'{i}nez}(2016)}]{Pozas-Kerstjens:2016}%
  \BibitemOpen
  \bibfield  {author} {\bibinfo {author} {\bibfnamefont {A.}~\bibnamefont
  {Pozas-Kerstjens}}\ and\ \bibinfo {author} {\bibfnamefont {E.}~\bibnamefont
  {Mart\'{i}n-Mart\'{i}nez}},\ }\href {\doibase 10.1103/PhysRevD.94.064074}
  {\bibfield  {journal} {\bibinfo  {journal} {Phys. Rev. D}\ }\textbf {\bibinfo
  {volume} {94}},\ \bibinfo {pages} {064074} (\bibinfo {year}
  {2016})}\BibitemShut {NoStop}%
\bibitem [{\citenamefont {Lifschytz}\ and\ \citenamefont
  {Ortiz}(1994)}]{Lifschytz1994}%
  \BibitemOpen
  \bibfield  {author} {\bibinfo {author} {\bibfnamefont {G.}~\bibnamefont
  {Lifschytz}}\ and\ \bibinfo {author} {\bibfnamefont {M.}~\bibnamefont
  {Ortiz}},\ }\href {\doibase 10.1103/PhysRevD.49.1929} {\bibfield  {journal}
  {\bibinfo  {journal} {Phys. Rev. D}\ }\textbf {\bibinfo {volume} {49}},\
  \bibinfo {pages} {1929} (\bibinfo {year} {1994})}\BibitemShut {NoStop}%
\bibitem [{\citenamefont {{Henderson, Laura}}(2021)}]{henderson:2021}%
  \BibitemOpen
  \bibfield  {author} {\bibinfo {author} {\bibnamefont {{Henderson, Laura}}},\
  }\emph {\bibinfo {title} {What can detectors detect?}},\ \href
  {http://hdl.handle.net/10012/16769} {Ph.D. thesis} (\bibinfo {year}
  {2021})\BibitemShut {NoStop}%
\end{thebibliography}%

\end{document}